\documentclass{aa}
\usepackage[utf8]{inputenc}
\usepackage{newunicodechar}
\newunicodechar{−}{\ensuremath{-}}
\newunicodechar{₂}{$_2$}
\usepackage{booktabs}
\usepackage{algorithm2e}
\usepackage{pdflscape}
\usepackage{array}
\usepackage{gensymb} 
\usepackage{amsmath}
\usepackage{adjustbox}
\usepackage{lipsum} 
\usepackage{rotating}
\usepackage{float} 
\usepackage{graphicx}
\usepackage[english]{babel}
\usepackage[pagebackref]{hyperref}
\hypersetup{colorlinks=true, citecolor=blue,linkcolor=black,urlcolor=blue}
\usepackage{natbib} 
\usepackage{subfig}
\usepackage{txfonts}
\usepackage{xcolor} 
\usepackage{comment}

\begin{document}  

   \title{Resolved Dust-Gas-Metallicity relations in nearby spiral galaxies}
   \subtitle{}
\author{
    Vidhi Tailor\inst{1,2}
    \and Viviana Casasola\inst{1}
    \and Francesco Calura\inst{3}
    \and Francesca Pozzi\inst{2,3}
    \and Jacopo Fritz\inst{4,1}
    \and Maritza Lara-López\inst{5,6}
    \and Marco Palla\inst{7}
    \and Simone Bianchi\inst{7}
    \and Santiago Álvarez-Córdoba\inst{4}
    \and Matteo Bonato\inst{1,8}
    \and María Emilia De Rossi\inst{9,10}
    \and Sami Dib\inst{11}
    \and Luis E. Garduño\inst{5,6}
    \and Andrea Giannetti\inst{1}
    \and Omar López Cruz\inst{12}
    \and Marina M. Puebla\inst{5,6}
    \and Evangelos D. Paspaliaris\inst{7}
    \and Leonid S. Pilyugin\inst{13,14}
    \and Guillermo Valé\inst{5,6}
    \and Mabel Valerdi\inst{15}
}

\institute{
    INAF -- Istituto di Radioastronomia, Via Gobetti 101, I-40129 Bologna, Italy
    \email{vidhiritesh.tailor@unibo.it}
    \and
    Dipartimento di Fisica e Astronomia, Alma Mater Studiorum Università di Bologna, Via Piero Gobetti 93/2, I-40129 Bologna, Italy
    \and
    INAF -- Osservatorio di Astrofisica e Scienza dello Spazio di Bologna, Via Gobetti 93/3, I-40129 Bologna, Italy
    \and
    Instituto de Radioastronomía y Astrofísica, UNAM, Campus Morelia, A.P. 3-72, Morelia 58089, Mexico
    \and
    Departamento de Física de la Tierra y Astrofísica, Fac. de C.C. Físicas, Universidad Complutense de Madrid, E-28040 Madrid, Spain
    \and
    Instituto de Física de Partículas y del Cosmos, IPARCOS, Fac. C.C. Físicas, Universidad Complutense de Madrid, E-28040 Madrid, Spain
    \and
    INAF -- Osservatorio Astrofisico di Arcetri, Largo E. Fermi 5, I-50125 Firenze, Italy
    \and
    INAF -- Italian ALMA Regional Centre, Via Gobetti 101, I-40129 Bologna, Italy
    \and
    Universidad de Buenos Aires, Facultad de Ciencias Exactas y Naturales y Ciclo Básico Común, Buenos Aires, Argentina
    \and
    CONICET--Universidad de Buenos Aires, Instituto de Astronomía y Física del Espacio (IAFE), Buenos Aires, Argentina
    \and
    Max-Planck-Institut für Astronomie, Königstuhl 17, D-69117 Heidelberg, Germany
    \and
    Instituto Nacional de Astrofísica, Óptica y Electrónica (INAOE), 
    Luis E. Erro No.~1, Sta.\ Ma.\ Tonantzintla, Puebla, C.P.~72840, Mexico
    \and
    Institute of Theoretical Physics and Astronomy, Vilnius University, Sauletekio av. 3, 10257, Vilnius, Lithuania
    \and
    Main Astronomical Observatory, National Academy of Sciences of Ukraine, 27 Akademika Zabolotnoho St., 03143, Kyiv, Ukraine
    \and
    Departamento de Astronomía, Universidad de Chile, Camino del Observatorio 1515, Las Condes, Santiago, Chile
}

   \date{\today}

\abstract
  {Understanding the interstellar medium (ISM) requires high-resolution, multi-component mapping to capture its complex physical structure.
  Nearby spiral galaxies, with their abundant and diverse ISM, provide an ideal laboratory for such a comprehensive analysis at sub-galactic scales.
  }
  {We investigate dust-to-gas (DGR) and dust-to-metal (DMR) ratios as a functions of gas-phase metallicity ($Z$), on spatial scales ranging from 0.6 to 2.3 kpc, in a sample of 10 nearby spiral galaxies, spanning more than an order of magnitude in stellar mass ($9.7 \le \log(M_*/M_\odot) \leq 11.0$), star formation rate (SFR, $\sim$0.3--$3 \, M_\odot \, \rm yr^{-1}$) and metallicity ranging from $8.3 \lesssim 12 + \log(\rm O/H) \lesssim 8.8$. We explore how the DGR--$Z$ and DMR--$Z$ relations are shaped by the assumptions behind the CO-to-H$_2$ conversion factor ($\alpha_{\rm CO}$). 
  }
  {We homogeneously combine maps of dust, atomic gas, molecular gas, and metallicity. Motivated by the diversity in $L_{\rm CO(1-0)}$/SFR ratios and metallicity across our sample, we introduce a hybrid $\alpha_{\rm CO}$ prescription to distinguish between CO-bright and CO-dark regimes.
  The derived DGR--$Z$ and DMR--$Z$ relations are compared with other global and resolved observational results, and with the predictions of dust and chemical evolution models.}
  {Both DGR--$Z$ and DMR--$Z$ relations are dependent on the adopted $\alpha_{\rm CO}$ prescription, and no single $\alpha_{\rm CO}$ can reproduce the properties of the entire sample, motivating the use of a hybrid approach.
  The DGR increases with metallicity, spanning $\sim$1 dex across the sampled range;
  while the DMR remains approximately constant at $\log(\mathrm{DMR}) = -0.53 \pm 0.13$, 
  implying that $\sim$30$\%$ of metals are locked into dust grains. This flat behavior indicates an evolved dust phase where efficient ISM grain growth drives a saturation regime balancing dust formation and destruction.}
  {}

   \keywords{galaxies: evolution, galaxies: general, galaxies: ISM, galaxies: spiral, galaxies: star formation, ISM:  dust, extinction} 

   \maketitle
\begin{table*}[ht!]
    \caption{Main properties of the sample galaxies.}
    \centering
    \begin{tabular}{lccccccccl}
        \hline\\      
        Galaxy &  \(\alpha_{\rm J2000}\) &  \(\delta_{\rm J2000}\) & RC3 type & \(D_{25}\) &  Distance &  \textit{i} &  Nuclear & Resolution  & \(N_{pix}\)\\
        &  [\(^{\rm h}\) \(^{\rm m}\) \(^{\rm s}\)] &  [\(^{\circ}\) \(^{\prime}\) \(^{\prime\prime}\)] & &  [\(^{\prime}\)] & [Mpc] &  [\(^{\circ}\)] & activity  &[kpc] &\\
        (1) & (2) & (3) & (4) & (5) & (6) & (7) & (8) &(9)&(10)\\
        \hline
        NGC 2403        & 07 36 51.1 & +65 36 03 & SAB(s)cd      & 20.0 & 3.18            & 62.9  &H{\sc ii}\textsuperscript{(d)} & 0.6 & 981 \\
        NGC 925         & 02 27 16.5 & +33 34 44 & SAB(s)d       & 10.7 & 8.67             & 66.0  & H{\sc ii}\textsuperscript{(d)}  &1.5& 801\\
        NGC 6946        & 20 34 52.2 & +60 09 14 & SAB(rs)cd     & 11.5 & 4.51\textsuperscript{(a)} & 32.6  & H{\sc ii}\textsuperscript{(c)}& 0.8 &1467\\
        NGC 628  (M~74)      & 01 36 41.8 & +15 47 00& SA(s)c    & 10.0  & 8.83\textsuperscript{(b)}      & 7.0  & H{\sc ii}\textsuperscript{(c)}& 1.5&555\\
         NGC 4736 (M 94) & 12 50 53.0 & +41 07 13 & (R)SA(r)ab    & 7.8  & 4.59             & 41.4  & LINER\textsuperscript{(d)}& 0.8&134\\
        IC 342 & 03 46 48.5 & +68 05 47& SAB(rs)cd    & 20.0  & 3.39      & 31.0  & H{\sc ii}\textsuperscript{(d)}& 0.6&209\\
        NGC 5457 (M 101)& 14 03 12.6 & +54 20 57 & SAB(rs)cd     & 24.0 & 6.95           & 18.0  & H{\sc ii}\textsuperscript{(d)}& 1.2&803\\
        NGC 5194 (M 51) & 13 29 52.7 & +47 11 43 & SA(s)bc pec   & 13.8 & 7.55\textsuperscript{(b)} & 42.0  & Seyfert\textsuperscript{(d)}& 1.3&744\\
        NGC 5055 (M 63) & 13 15 49.2 & +42 01 45 & SA(rs)bc      & 11.8 & 8.99            & 59.0  &  LINER\textsuperscript{(d)}& 1.6&260\\
        NGC 3521        & 11 05 48.6 & -00 02 09 & SAB(rs)bc     & 8.3  & 13.24            & 72.7  & LINER\textsuperscript{(c)}&2.3& 152 \\
          
        \hline
    \end{tabular}
    \label{tab:sample}
    \vspace{2mm}
    \tablefoot{Main properties of the 10 galaxies analyzed in this work. The columns list the following: (1) Galaxy name; (2) Right ascension (\(\alpha_{\rm J2000}\)) and (3) Declination (\(\delta_{\rm J2000}\)) in J2000 coordinates; (4) RC3 classification type; (5) Optical diameter; (6) Galaxy distance; (7) Galaxy inclination angle; (8) Nuclear activity; (9) Physical resolution (FWHM beam, 36"); (10) \(N_{pix}\), number of valid pixels included in the analysis for each galaxy. The galaxies are arranged in order of increasing stellar mass  (see Table~\ref{tab:Integrated}). Distances are primarily taken from the Cosmicflows-3 database \citep{Tully_2016}, except where noted otherwise: \textsuperscript{(a)}~\cite{Dist_N6946} and \textsuperscript{(b)}~\cite{Dist_N5194_N628}. The nuclear activity classification is from: \textsuperscript{(c)}~\cite{Goulding_Class_c} and \textsuperscript{(d)}~\cite{Lem_cass_d}. Source references for other galaxy properties in C17.}    
\end{table*}

\section{Introduction}
The evolution of galaxies is regulated by several physical processes, among which the cycling of gas, metals, and dust within the interstellar medium (ISM) plays a key role.
Heavy elements produced by stars are returned to the ISM via stellar winds and supernova explosions, where they may remain in the gas phase or condense onto dust grains \citep[e.g.,][]{Dwek_1998}. 
Dust forms through metal condensation in the cool atmospheres of AGB stars \citep[e.g.,][]{Ferra_2006, Nanni_2013, Boyer_2025} and in the ejecta of supernovae (e.g., \citealt{Matsuura_2015, temim_2017, DeLooze_19}; see \citealt{Calura_26} for a review).
However, stellar sources alone cannot explain the large dust masses that are observed, leading to a widely accepted scenario in which a significant fraction of dust mass is built up through accretion of metals onto pre-existing grains within the dense ISM \citep{Dwek_1998,Calura_2008,Zhuko_2016,DeVis_2017,Popping_2017}.
The relative contribution of metals in the gas and dust phases therefore reflects the balance between dust production by stellar sources, grain growth in the ISM, and destruction processes \citep[e.g., supernova shocks,][]{Jones_1996,Slavin_2015,Priestley_2021}, providing information about the galactic dust cycle and the evolution of the ISM.
Observationally, we commonly study this balance through ratios such as the dust-to-gas ratio (DGR), dust-to-stellar ratio (DSR), and dust-to-metal ratio (DMR) \citep[e.g.,][]{RemyRuyer_2014,Calura_2017,Popping_2017, DeVis_2019,Galliano_2018,Casasola_2020,Casasola_2022}.

Early studies have established a tight correlation between the DGR and gas-phase metallicity (denoted by $Z$ or 12 + log(O/H) in this study), following an approximately linear scaling relation for nearby galaxies \citep[e.g.,][]{Draine_2007}, which suggests a roughly constant fraction of metals locked in dust. 
However, subsequent work extended this picture to lower metallicity systems and showed a significant deviation from linearity, with a steep decline in DGR at sub-solar metallicities.
\cite{RemyRuyer_2014} investigated 126 nearby galaxies spanning a broad metallicity range (\(\rm 12+log(O/H) \sim 7.1-9.1\)) and  demonstrated that the global DGR--$Z$ relation is more accurately described by a broken power law, with a steep decline below \(\rm 12 + log(O/H) \lesssim 8.0\), interpreted as a regime where dust production is dominated by stellar sources due to inefficient grain growth. 
In contrast, a later study of 466 late-type galaxies, by \cite{DeVis_2019}, in the frame of the Dustpedia collaboration \citep{Davies_2017}, suggested that the DGR--Z relation could instead be described by a single continuous power law, implying a more gradual transition in dust build-up. 
More recent work combining multi-wavelength datasets has continued to debate the exact shape and universality of this relation, highlighting the role of sample selection, calibration systematics, and ISM conditions \citep[e.g.,][]{Cortese_2016,Casasola_2020,Galliano_2021}.

In recent years, spatially resolved studies have enabled the investigation of these relations on (sub-)kpc scales in nearby galaxies. 
\cite{Sandstrom_2013} analyzed the DGR across 26 nearby star-forming galaxies, finding that it correlates linearly with metallicity on kpc scales.
However, subsequent studies of individual systems \citep[e.g., NGC~628 (M~74), NGC~5457 (M~101), M~33;][]{Relano_2018, Vilchez_2019} revealed a steepening of the DGR--$Z$ relation below $12 + \log(\text{O/H}) \approx 8.4$, with \cite{Vilchez_2019} measuring a slope of \(\sim 1.33\) in this low-metallicity regime for NGC~5457.
This superlinear trend is further supported by the findings of \cite{Casasola_2022}, who reported a slope of 1.2 -- 1.3 in the log(DGR) vs. \(\rm 12 + \log (O/H)\) relation for metallicities ranging from 8.2 to 8.8.
While these studies provide valuable insights, they often rely on assumed metallicity gradients derived from a limited spatial coverage rather than direct measurements and are limited in sample size.

Integral Field Unit (IFU) surveys now facilitate direct, spatially resolved metallicity measurements across galactic disks. 
Using data from the TYPHOON survey, \cite{Park_2024} found that the spatially resolved DGR--$Z$ relation is better described by a broken power law, characterized by significant scatter at intermediate metallicities ($8.0 < 12 + \log(\text{O/H}) < 8.3$) and an anomalously high DGR in the dwarf galaxy Sextans~A. 
These results, along with other resolved studies \citep[e.g.,][]{Roman_Duval_2021, Clark_2023}, suggest that dust content is regulated not only by metallicity but also by local ISM conditions and the multi-phase structure of the gas.

A more direct probe of these processes is provided by the DMR, which traces the efficiency of dust growth. 
In the Milky Way (MW), depletion studies indicate that $\sim 20$--$50\%$ of metals are incorporated into the dust phase, with this fraction increasing as a function of gas density \citep[e.g.,][]{Jenkins_2009}. 
Such environmental variations are also observed with high spatial resolution in nearby systems such as the Magellanic Clouds \citep[e.g.,][]{Roman_2014}, confirming that the DMR is sensitive to local ISM environment.
On larger scales, global extragalactic studies show that the DMR generally increases with metallicity before saturating in metal-rich systems \citep[e.g.,][]{DeVis_2019, Galliano_2021}. 
This is consistent with the resolved findings of \cite{Casasola_2022}, who reported a lack of a significant trend between DMR and metallicity in their sample \citep[see also][for an unresolved sample at a similar metallicity range]{DeLooze_2020}. Oxygen depletion studied in the H{\sc ii} regions by \citet{Peimbert_2010} suggests that a nearly constant fraction of oxygen (\(\sim 0.09 - 0.11\) dex) is incorporated into dust grains accross the metallicity range \(\rm 7.3 < 12 + log(O/H) < 8.8\).
This suggests that, in the metallicity regime typical of many nearby spirals \(\rm{(12 + log(O/H) \geq 8.1-8.2)}\), the DMR may already have reached a plateau, potentially indicating that the critical threshold for efficient grain growth lies in lower metallicities. 

This behavior is explained by dust evolution models, which track the balance between dust sources and sinks \citep[e.g.,][]{Asano_2013,Z_2014,Feldmann_2015,Popping_2017}.
A key prediction of these models is the existence of a "critical metallicity" above which grain growth becomes efficient, rapidly increasing dust content and driving the DMR towards a near-constant value.
Below this threshold, stellar sources (AGBs, SNe) should be the main dust producers, implying lower DMR ratios. 
These models differ significantly in their assumptions, reflecting both uncertainties in the underlying physical processes and the lack of strong observational constraints, particularly at low metallicity. Constraining the efficiency of grain growth and the other dust processes is therefore essential for understanding the origin and evolution of dust in galaxies.

In this work, we investigate the spatially resolved DGR--Z and DMR--Z relations across a sample of 10 nearby late-type galaxies, using observations that sample physical scales ranging from $\sim 0.6$ to $2.3$~kpc. 
We also explore how these relations are influenced by different prescriptions for the CO-to-$\rm H_2$ conversion factor ($\alpha_{\rm CO}$).
Unlike integrated measurements, spatially resolved observations provide a more detailed view of ISM physics. 
These data offer stringent constraints for evolutionary models and establish an important 
local benchmark for interpreting global scaling relations and high-redshift systems.

The paper is organized as follows. We describe the galaxy sample and our dataset in Section~\ref{sec:sample}. In Section~\ref{sec:methods}, we present the methods used to measure ISM parameters (gas-phase metallicity, \(\mathrm{HI}\) and \(\rm H_2\) masses) for resolved regions and discuss the various $\alpha_{\rm CO}$ prescriptions used in the study. We present our DGR--$Z$ and DMR--$Z$ relations for different \(\alpha_{\rm CO}\) in Section~\ref{sec:results}, and compare our findings to previous studies and galaxy evolution models in Section~\ref{sec:discussion}. We then summarize our conclusions in Section~\ref{sec:conclusions}.

\section{Galaxy sample and dataset}
\label{sec:sample}

\begin{table*}[ht!]
\caption{Integrated physical properties of the sample galaxies.}
\centering
\begin{tabular}{lccccccc}
\hline
Galaxy & $\log (M_\star)$& $L_{\rm CO (1-0)}$  & $\log (M_{\mathrm{H\,I}})$& $\log (M_{\rm dust})$& SFR & $\rm 12 + log(O/H)$\\
 &  [$\rm M_{\odot}$]& [$\rm L_{\odot}$] &  [$\rm M_{\odot}$]&  [$\rm M_{\odot}$]& [$\rm M_\odot\,{\rm yr}^{-1}$] & \\
\hline
NGC\,2403 & $9.72 \pm 0.22$ &$9.02 \times10^2$& $8.92 \pm 0.04$ & $6.63 \pm 0.03$ & $0.33 \pm 0.19$ & 8.36 \\
NGC\,925 & $9.82 \pm 0.22$ &$1.87 \times10^3$&  $9.19 \pm 0.04$ & $6.91 \pm 0.03$ & $0.36 \pm 0.18$ & 8.25 \\
NGC\,6946 & $10.27 \pm 0.17$ &$1.51 \times10^4$&  $9.08 \pm 0.04$ & $7.09 \pm 0.03$ & $1.42 \pm 0.45$ & 8.58 \\
NGC\,628 & $10.30 \pm 0.21$ &$1.23 \times10^4$  & $9.57 \pm 0.04$ & $7.20 \pm 0.03$ & $1.08 \pm 0.54$ & 8.48\\
NGC\,4736 & $10.34 \pm 0.15$ &$3.16 \times10^3$& $8.17 \pm 0.04$ & $6.34 \pm 0.03$ & $0.48 \pm 0.24$ & 8.60 \\
IC\,342 & $10.61 \pm 0.08$ &$3.70 \times10^4$  & $9.27 \pm 0.04$ & $7.26 \pm 0.03$ & $1.66 \pm 0.72$ & 8.53 \\
NGC\,5457 & $10.66 \pm 0.12$ &$2.19 \times10^4$& $10.21 \pm 0.04$ & $7.51 \pm 0.03$ & $2.88 \pm 1.41$ & 8.50 \\
NGC\,5194 & $10.72 \pm 0.12$ &$3.54 \times10^4$& $9.30 \pm 0.04$ & $7.42 \pm 0.03$ & $3.01 \pm 0.81$ & 8.62 \\
NGC\,5055 & $10.79 \pm 0.17$ &$2.38\times10^4$ & $9.35 \pm 0.04$ & $7.56 \pm 0.03$ & $1.85 \pm 0.96$ & 8.61 \\
NGC\,3521 & $10.98 \pm 0.18$ &$3.22 \times10^4$  & $9.34 \pm 0.04$ & $7.61 \pm 0.03$ & $3.18 \pm 0.22$ & 8.71 \\
\hline
\end{tabular}
\label{tab:Integrated}
\vspace{2mm}
\tablefoot{
Integrated quantities within the optical radius $R_{25}$. Uncertainties on \(M_{\rm dust}\) are estimated via Monte Carlo bootstrapping, assuming for each pixel a surface brightness uncertainty equal to the quadrature sum of the photometric, background, and calibration errors in each band. Uncertainties on \(M_*\) and \(\rm SFR\) are computed as the quadrature sum of the background and calibration error. The uncertainty on \(M_{\mathrm{HI}}\) is computed as the quadrature sum of the \(\mathrm{HI}\) flux and distance uncertainties, assuming a 10\% distance error, which usually dominates the total error. The final column lists the integrated (global) gas-phase metallicity of each galaxy, expressed as \(\rm 12 + \log(O/H)\); the methodology used to derive these values is described in Section~\ref{sec:metallicity}. } 
\end{table*}

\begin{figure}[ht]
\centering

\setlength{\tabcolsep}{2pt}
\renewcommand{\arraystretch}{0}

\begin{tabular}{cc}

\includegraphics[width=0.48\linewidth]{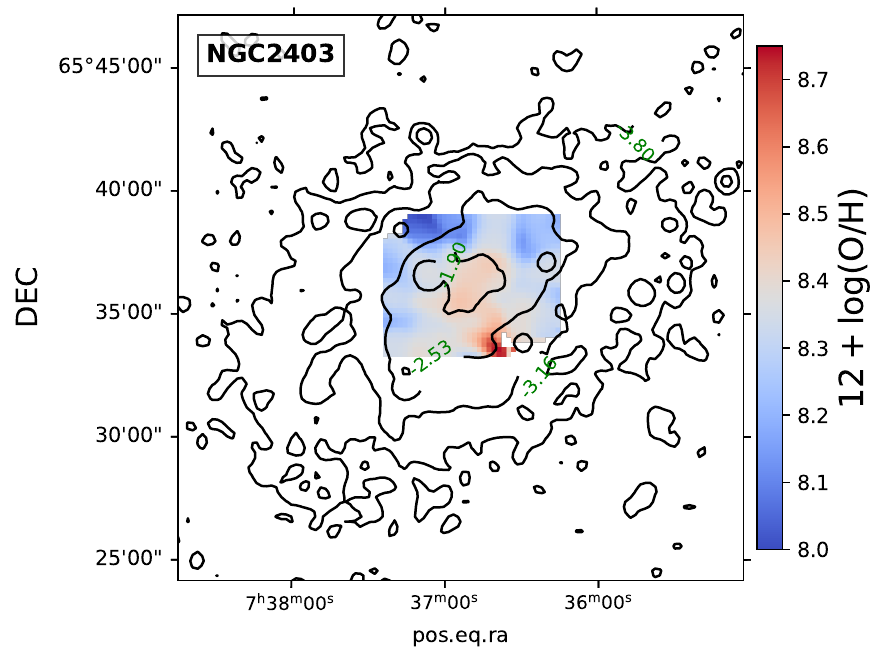} &
\includegraphics[width=0.45\linewidth]{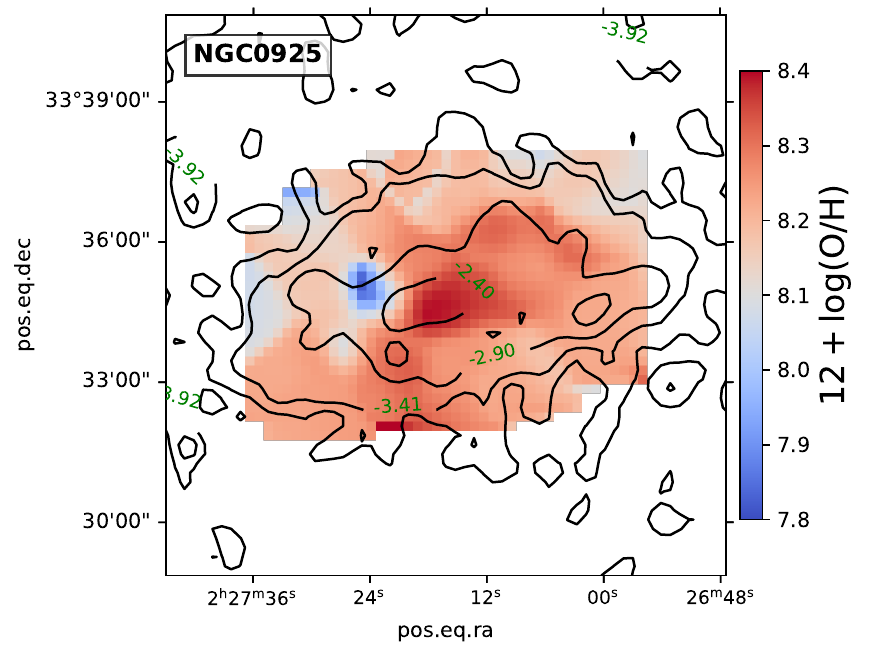} \\

\includegraphics[width=0.48\linewidth]{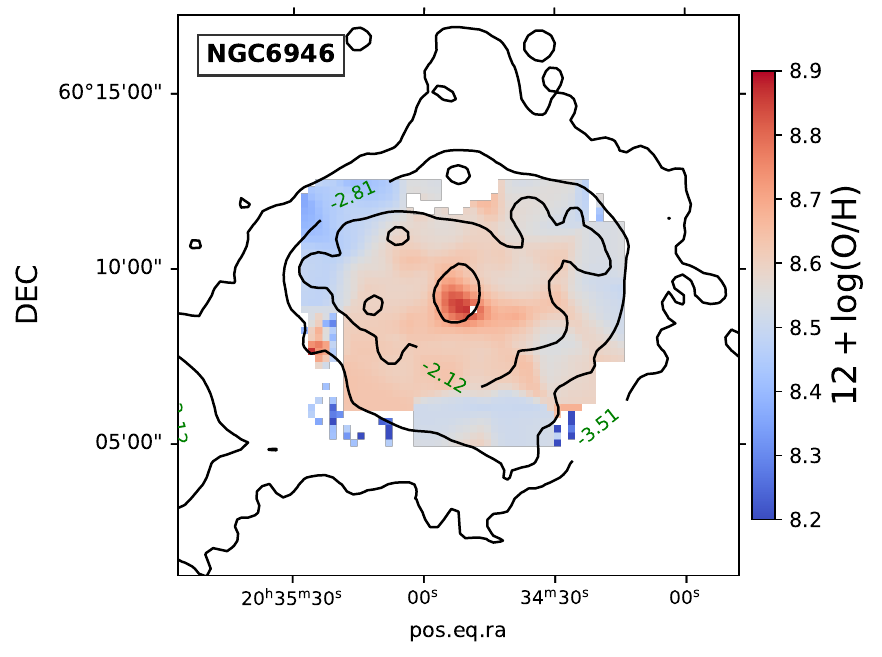} &
\includegraphics[width=0.45\linewidth]{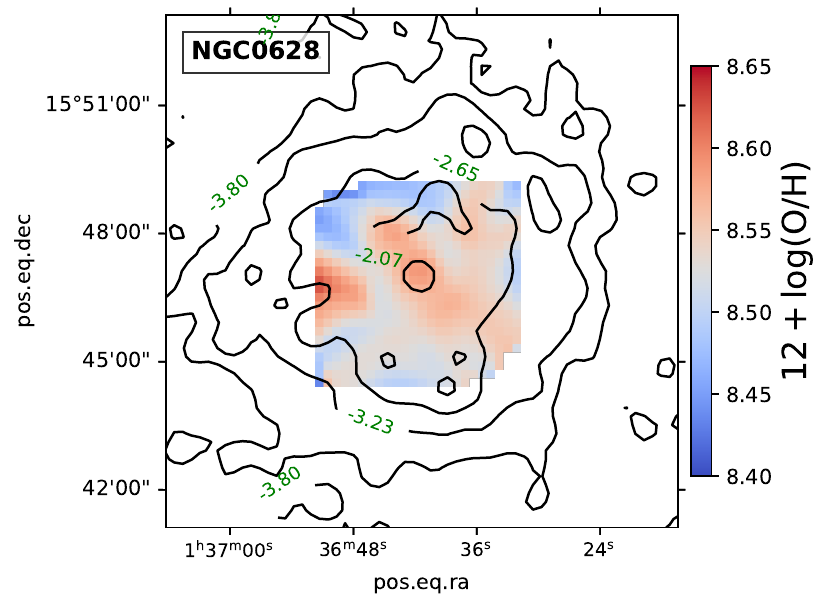} \\

\includegraphics[width=0.48\linewidth]{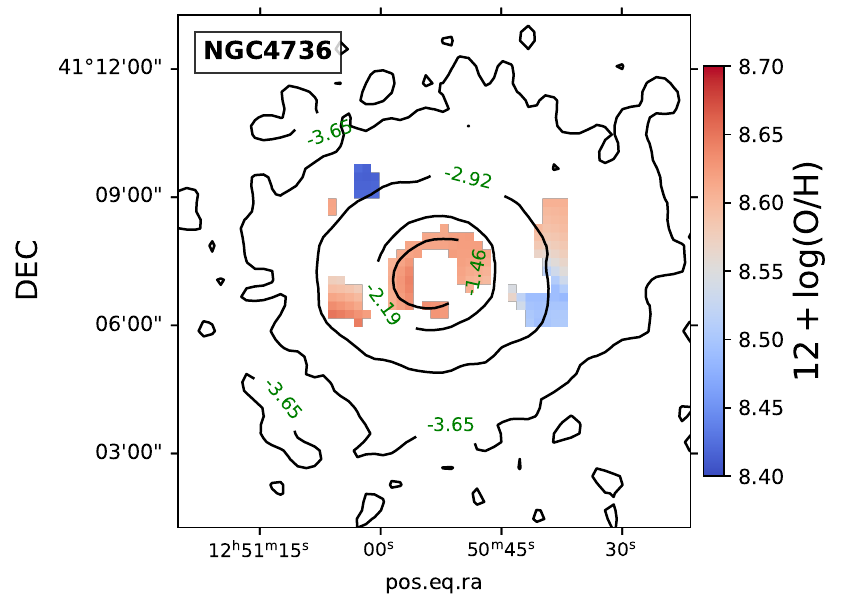} &
\includegraphics[width=0.45\linewidth]{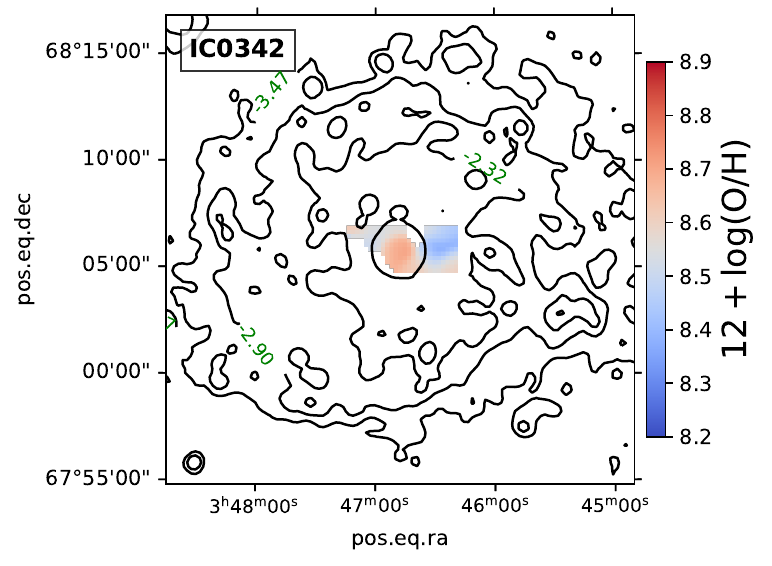} \\

\includegraphics[width=0.48\linewidth]{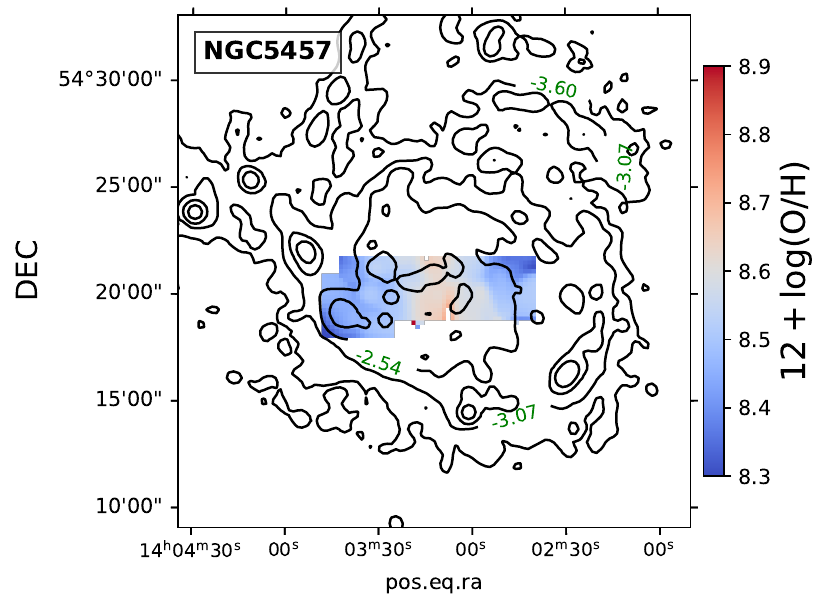} &
\includegraphics[width=0.45\linewidth]{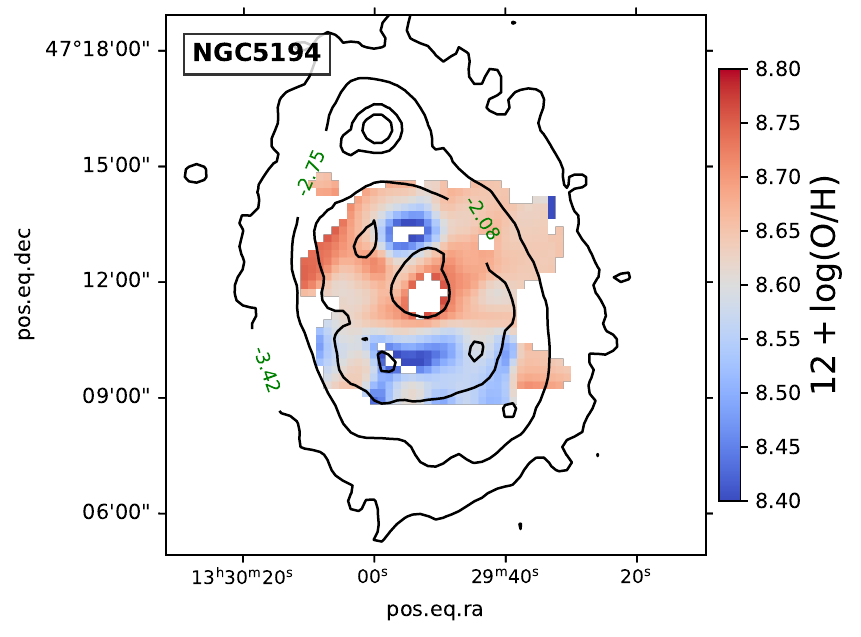} \\

\includegraphics[width=0.48\linewidth]{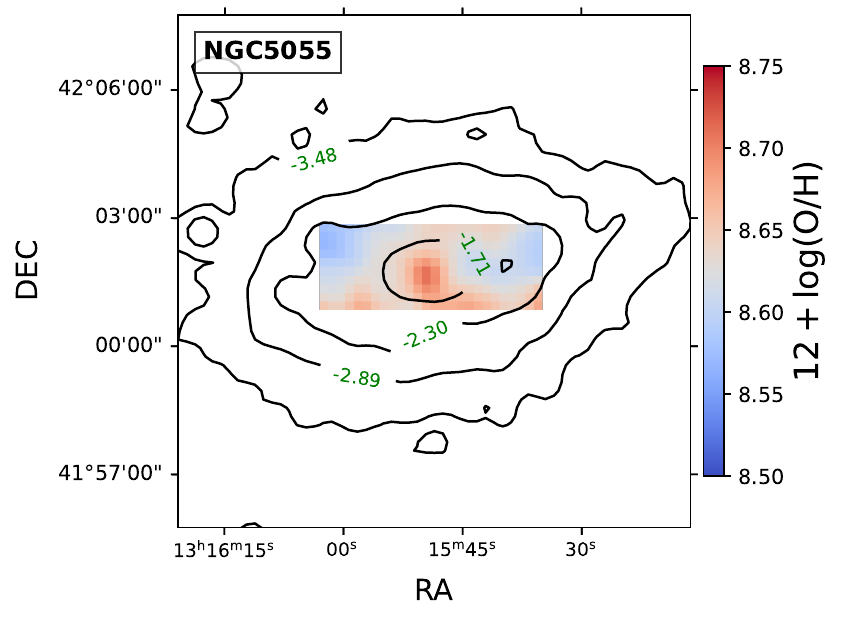} &
\includegraphics[width=0.45\linewidth]{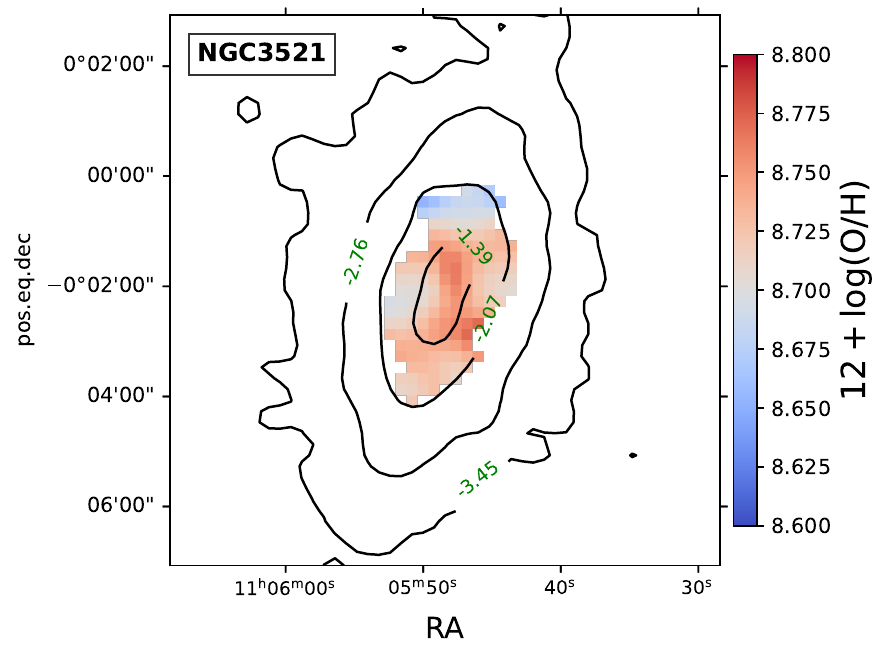}

\end{tabular}
\caption{Gas-phase metallicity (characterized by $12+\log(\mathrm{O/H})$; see Sect.~\ref{sec:metallicity}) maps for our sample galaxies with the superposition of (black) contours of the $\Sigma_{\mathrm{SFR}}$ distribution (in units of $M_{\odot} \, \mathrm{yr}^{-1} \, \mathrm{kpc}^{-2}$; see Sect.~\ref{sec:dust_stellar_spatial coverage}). For each galaxy, four logarithmically spaced $\Sigma_{\mathrm{SFR}}$ contours are shown, spanning from the 50th to the 99th percentile of its $\Sigma_{\mathrm{SFR}}$ distribution. The green labels indicate the corresponding values of \( \log_{10} (\Sigma_\mathrm{SFR})/[M_{\odot} \, \mathrm{yr}^{-1} \, \mathrm{kpc}^{-2}]\); the contour levels are therefore determined independently for each galaxy.}
\label{fig:metallicity_coverage}
\end{figure}

Our galaxy sample has been selected from the intersection between the DustPedia sub-sample \citep[][hereafter C17]{Casasola_2017} and the Metal-THINGS survey sample \citep{Lara_Lopez_2021}, supplemented by the galaxy NGC~628 from the PHANGS-MUSE survey \citep{Emsellem_22}. 
To achieve a multi-phase view of the ISM on (sub-) kpc scales in nearby galaxies, we combined several spatially resolved datasets covering dust, gas, stars, and star formation rate (SFR).
The distribution of these latter quantities is obtained following the DustPedia analysis described in C17 and \cite{Casasola_2022}. 

The DustPedia project \citep{Davies_2017} provides a comprehensive multi-wavelength characterization of 875 nearby galaxies (within 40~Mpc) observed with \textit{Herschel}. 
From this parent sample, C17 selected 18 large face-on spirals that are well-resolved in the submillimeter, with the galaxy extent measured from Herschel/SPIRE 250~$\mu$m images at the 12 mJy/beam isophote satisfying \(D_{\rm sub-mm} \geq 9'\).
In this work, the atomic ($\text{HI}$) and molecular (CO) gas maps are produced using observations from the THINGS \citep{Walter_2008} and HERACLES \citep{Leroy_2009} surveys, respectively (except for IC~342, where VLA and NRO 45 \,m data are used; \citealt{Crosthwaite_2001, Kuno_2007}). 
The gas-phase metallicities are then integrated into this dataset, derived primarily from the Metal-THINGS survey \citep{Lara_Lopez_2021}. 
Metal-THINGS provides optical integral-field spectroscopy (IFS) for 25 galaxies using the George and Cynthia Mitchell Spectrograph \citep[GCMS, formerly VIRUS-P;][]{Hill_2008} on the 2.7\,m Harlan J. Smith Telescope at McDonald Observatory. The original Metal-THINGS Raw Stacked Spectra (RSS) were converted into data cubes by convolving them using a Gaussian with FWHM = 4.16\arcsec \citep[similar to][]{Garduno_2023}, establishing a pixel size of 4.16\arcsec, which were used to extract emission-line fluxes and derive resolved metallicity maps across the galaxy disks. 
For NGC~628, we utilize MUSE observations, processed to mimic the Metal-THINGS data by spatially binning to $4\arcsec$ pixels and degrading the spectral resolution from $2.6$ to $5.4\,\AA$ (Full Width at Half Maximum,  FWHM), and truncating the spectra at wavelengths longer than $\sim$7000~$\AA$.
We refer the reader to the aforementioned original papers for a more detailed description of the observation strategies and data reduction procedures for each survey.

The final sample, resulting from the cross-matching of these multi-wavelength datasets, consists of 10 nearby late-type galaxies whose main properties are listed in Table~\ref{tab:sample}. 
Their distances range from $\sim$3.2 to $\sim$13.2~Mpc and their Hubble types from Sab to Sd. 
These systems are characterized by large angular sizes ($D_{25} \ge 7.8'$ \footnote{$D_{25}$ is the major-axis diameter at which the $B$-band surface brightness reaches 25 mag arcsec$^{-2}$. 
We adopt $R_{25} = D_{25}/2$.}) and low-to-moderate inclinations (up to $\sim$70$\degree$), making them ideal targets for spatially resolved studies of the ISM. 
The sample includes both unbarred and weakly barred spirals, with nuclear classifications ranging from H{\sc ii} regions to LINERs and Seyfert galaxies. 
Physically, the sample spans slightly more than one order of magnitude in stellar mass ($9.7 \leq \log (M_*/M_{\odot}) \leq 11.0$) and covers a global metallicity range of $8.3 \leq 12 + \log(\rm O/H) \leq 8.8$. 
The SFR ranges from $\sim$0.3 to $3~M_{\odot}\,{\rm yr}^{-1}$, placing these galaxies along the star-forming main sequence of the local Universe \citep[e.g.,][]{CanoDiaz_2016}. 
These physical properties, together with the integrated ISM parameters, namely the CO luminosity ($L_{\rm CO(1-0)}$), the atomic gas mass ($M_{\rm HI}$), and the dust mass ($M_{\rm dust}$), are listed in Table~\ref{tab:Integrated}. They represent the integrated counterparts of the resolved maps described in the subsequent sections.

Figure~\ref{fig:metallicity_coverage} shows the spatial distribution and coverage of the gas-phase metallicity measurements across the galaxy disks. For each system, we present the metallicity map derived from the available spectroscopic data, with contours of the star formation rate surface density ($\Sigma_{\rm SFR}$; see Sect.~\ref{sec:dust_stellar_spatial coverage}) overlaid to indicate the extent of the star-forming disk and provide a visual reference for the spatial coverage of the resolved analysis.
In most cases, the metallicity measurements extend over a substantial radial fraction of the star-forming disk, enabling a robust resolved analysis of the DGR--$Z$ and DMR--$Z$ relations. 
While the spatial coverage for NGC~4736 is more limited, we include it in the analysis wherever valid measurements are available.
The use of 2D metallicity maps rather than radial gradients extrapolated from integrated measurements is a key strength of our dataset and study.
For instance, while \citet{Casasola_2022} previously investigated the DGR--$Z$ and DMR--$Z$ relations across our sample, their analysis relied on metallicity maps reconstructed from radial gradients. 
In contrast, our approach utilizes "real" and spatially complete metallicity distributions, providing a more accurate representation of the local ISM enrichment and its impact on the explored scaling relations.

To quantify the representativeness of our analysis, we compute the fraction of the star-forming disk covered by our metallicity measurements. For most galaxies in our sample, reliable gas-phase metallicities are recovered over $\sim$73\%–100\% of the spaxels with detectable emission line measurements, with NGC~4736 showing the lowest coverage at $\sim$30\%. 
To perform a consistent spatially resolved analysis, we process each dataset to derive the physical quantities necessary for our study. 
In the following subsections, we describe the measurement of emission-line fluxes from the IFS cubes, the derivation of dust, stellar, gas, and SFR surface densities, and the procedure used to homogenize the different datasets to a common spatial resolution and grid.

\subsection{Emission lines measurement}
\label{sec:metalthings_data}

We use reduced IFS data cubes from the Metal-THINGS survey for each galaxy. Emission-line fluxes are extracted on a spaxel-by-spaxel basis following the procedure described below.
The spectra in each cube are analyzed with the fitting tool {\sc sinopsis} \citep[see][for details]{Fritz_2007, Fritz_17}, which reproduces the observed spectra using a combination of simple stellar population (SSP) models. The code setup used for the fitting is described in \cite{AlvarezCordoba2026}. While {\sc sinopsis} includes SSP models with pre-computed emission-line intensities, it also generates purely stellar spectra. These stellar models are subtracted from the observed spectra to produce gas “emission-only” data cubes, containing only the nebular emission component and accounting for the underlying stellar absorption. These emission-only cubes are then processed with a custom routine (described in Osorio-Quiroga et al. in prep.) that fits Gaussian functions to the emission lines in order to derive their redshifts, widths, and fluxes. 
We assess the consistency and reliability of this procedure by comparing the derived metallicities with those reported by \cite{Vale_25}, who analyzed the same data using an independent method to measure emission-line intensities. We find excellent agreement between the two approaches.

The final data products include extinction-corrected emission-line fluxes and associated uncertainties for H$\beta$, [\ion{O}{iii}]~$\lambda4959, 5007$, H$\alpha$, [\ion{N}{ii}]~$\lambda6548 , 6583$, and [\ion{S}{ii}]~$\lambda6716,6731$, where the extinction correction is derived from the Balmer decrement with \(\rm H\alpha/H\beta = 2.86\) \citep{Osterr_2006} and adopting the \cite{Cardelli_1989} extinction curve with \(R_V = 3.1\), along with measurements of the local continuum noise.
We note that the observed \(\rm H\alpha/H\beta\) ratio does not necessarily trace the total dust distribution within a galaxy. We find that its correlation with $\Sigma_{\rm dust}$ is strongest in nearly face-on galaxies and becomes substantially weaker in more inclined systems, likely reflecting the increasing complexity of dust--gas mixing along the line of sight and the limited sensitivity of optical observations to probe heavily obscured regions.
These emission-line maps are used to derive the gas-phase metallicities described in Section~\ref{sec:metallicity}.

\subsection{Surface density maps of dust mass, stellar mass, and SFR}
\label{sec:dust_stellar_spatial coverage}

We adopted the surface density maps of dust mass ($\Sigma_{\rm dust}$), stellar mass ($\Sigma_{\rm M_*}$), and star formation rate ($\Sigma_{\rm SFR}$) presented in C17 for the galaxies in our sample.
The $\Sigma_{\rm dust}$ is derived by fitting far-infrared SEDs with the THEMIS dust model \citep{Jones_2013,K_hler_2014,Jones_2017}, assuming dust grains heated by a single diffuse interstellar radiation field (ISRF) scaled by an intensity factor $U$, where $U=1$ corresponds to the Solar neighborhood ISRF. 

Within THEMIS, the dust is described as a mixture of amorphous carbonaceous and silicate grains, with the two populations physically coupled in core-mantle structures \citep{Jones_2013,Ysard_2015}. 
Both grain families span sizes from nanometer particles upto a few micrometers and are characterised by size-dependent optical properties. In the diffused ISM, silicate grains are further coated by thin carbonaceous mantles and incorporate Fe and FeS nano-inclusions, which modify their optical properties.

The fitting was performed using \textit{Herschel} bands at $\lambda \geq 160\,\mu$m (PACS 160 $\mu$m; SPIRE 250, 350, and 500 $\mu$m). When available, PACS 100 $\mu$m data were included, while PACS 70 $\mu$m fluxes were treated as upper limits to minimize contamination from stochastically heated small grains \citep{R_my_Ruyer_2013}. 

The $\Sigma_{\rm M_*}$ was derived from the images of IRAC \(3.6 \, \mu\)m and \(4.5 \, \mu\)m, following \cite{Querejeta_2015}. 
This method uses independent component analysis \citep[ICA;][]{Meidt_2011} to separate the dominant old stellar light from non-stellar contributions, such as polycyclic aromatic hydrocarbon (PAH) emission and the hot dust continuum \citep{Flagey_2006}. The stellar mass estimates were derived assuming a Chabrier initial mass function \citep[IMF,][]{Chabrier_2003}. For more details on implementing ICA to obtain stellar mass, refer to \cite{Querejeta_2015}.

The $\Sigma_{\rm SFR}$ was obtained by combining GALEX-FUV and WISE \(22 \, \mu\)m data using the calibration of \cite{Bigiel_2008}:
\begin{equation}
    \Sigma_{\text{SFR}} = 3.2 \times 10^{-3} \times I_{22} + 8.1 \times 10^{-2} \times I_{\text{FUV}},
\end{equation}
\noindent
where \(\Sigma_{\text{SFR}}\) is in units of \(M_{\odot} \, \text{yr}^{-1} \, \text{kpc}^{-2}\), 
and \(I_{22}\) and \(I_{\text{FUV}}\) are the \(22 \, \mu\)m and FUV intensities in units of \(\text{MJy} \, \text{sr}^{-1}\), respectively. This calibration is based on the IMF from \cite{Calzetti_2007}, taken from the default in Starburst99.
Such hybrid tracers (e.g., FUV+IR or $\rm H_{\alpha}$+IR) are widely used to estimate SFR in nearby galaxies \citep{Kennicutt_2007,Rahman_2011,Muraoka_2019,Yajima_2021,Casasola_2022}, as they also account for star formation that is obscured by dust and therefore not directly traced by FUV or \(\rm H_\alpha\) emission alone.

A detailed description of the derivation procedures and uncertainty estimates for all maps is provided in C17.

\subsection{$\mathrm{HI}$ and CO data}
\label{sec: gas_data}

We derived gas surface densities by using the \(\mathrm{HI}-21\) cm emission line to trace the atomic component and the \(\rm ^{12}CO\) emission lines to trace the molecular component, respectively.
The \(\mathrm{HI}\) - 21 cm data were obtained from "The HI Nearby Galaxy Survey"  \citep[THINGS;][]{Walter_2008}, which provides integrated VLA 21 cm intensity (moment 0) maps for 34 nearby galaxies at an angular resolution of $6''$, in units of  \(\rm \text{Jy}\,\text{beam}^{-1}\,m\,s^{-1}\). We used the natural weighted HI maps.
For IC~342 we used the \(\mathrm{HI} - 21\) cm map available from the NED catalog in units of  \(\rm \text{Jy}\,\text{beam}^{-1}\,m\,s^{-1}\) and obtained with the NRAO-VLA at an angular resolution of $38''$. 

The molecular gas data were derived using "The HERA CO-Line Extragalactic Survey" \citep[HERACLES,][]{Leroy_2009}, which provides \(\rm^{12}CO\, (2-1)\) integrated intensity maps obtained with the IRAM 30 m telescope at an angular resolution of $11''$ for 48 nearby galaxies. From the HERACLES archive, we extracted the reduced integrated intensity maps and associated uncertainties, in units of \(\rm K\,km\,s^{-1}\). 
Since IC~342 was not observed as part of HERACLES, we utilize the \(\rm ^{12}CO\, (1-0)\) line intensity map from \textit{“Nobeyama CO Atlas of Nearby Spiral Galaxies”} \citep{Kuno_2007} in units of \(\rm K\,km\,s^{-1}\), observed with the NRO 45 m telescope at an angular resolution of $15''$.

The derivation of gas surface densities and the adopted CO-to-\(\rm H_2\) conversion factors are described in Section~\ref{sec:method_gas}.

\subsection{Spatial resolution, convolution, and regridding}
\label{sec:resolution}

To enable a consistent, pixel-by-pixel comparison between the different ISM phases and stellar properties, all datasets were matched to a common angular resolution and spatial grid. 
The limiting resolution is set by the \textit{Herschel} / SPIRE $500\,\mu$m beam, with a FWHM of $36\arcsec$. 
Accordingly, the $\Sigma_{\rm dust}$, $\Sigma_{*}$, and $\Sigma_{\rm SFR}$ were adopted from \citet{Casasola_2017, Casasola_2022} at this common $36\arcsec$ resolution. 
For datasets with higher native resolution, the convolution to the common $36\arcsec$ beam was performed using Gaussian PSF matching kernels following \citet{Aniano_2011}.
An exception was made for IC~342, where the native resolution of the \(\mathrm{HI}\) data is $38\arcsec$; since this is comparable to our target resolution, no additional convolution was applied.
All maps were corrected for inclination, \textit{i}.

For the metallicity analysis, each emission-line map was convolved independently prior to computing the line ratios. 
This procedure ensures that the metallicity is derived from surface brightness distributions matched at the same spatial scale, preserving the consistency of the resolved physical properties. 

Finally, all maps were resampled onto a common $12\arcsec$ pixel grid, matching the WCS (World Coordinate System) of the $\Sigma_{\rm dust}$ maps.
At the distances of galaxies in our sample, the $36\arcsec$ angular resolution corresponds to physical scales ranging from $\sim$0.6 to 2.3~kpc, which defines the physical scale for our resolved analysis. To test for resolution effects, we repeated the analysis after homogenizing all maps to a common physical resolution of $2.3$~kpc. We find no significant changes in the derived quantities or in the resulting scaling relations, suggesting that variations in spatial resolution do not bias our results.

\section{Methodology}
\label{sec:methods}
\subsection{Gas-phase metallicity}
\label{sec:metallicity}

Gas-phase metallicity was derived from spatially resolved emission-line measurements provided by the Metal-THINGS survey.
Oxygen abundances were estimated using strong nebular emission lines tracing $\rm H\,\textsc{ii}$ regions across the galaxy disk. The [\ion{O}{iii}]~$\lambda4959$ and [\ion{N}{ii}]~$\lambda6548$ transitions are intrinsically weaker than their respective doublets at $\lambda5007$ and $\lambda6583$. We therefore assume a fixed flux ratio of $F_{5007}/F_{4959} \approx 3$ and $F_{6583}/F_{6548} \approx 3$ \citep[e.g.,][]{Storey_2000,Doj_2022} to maximize the number of usable spaxels and mitigate the impact of their low signal-to-noise ratio (S/N).
We require a S/N greater than 3 for all emission lines used for the metallicity calibration (H$\beta$, [\ion{O}{iii}]~$\lambda5007$, H$\alpha$, [\ion{N}{ii}]~$\lambda6583$, and [\ion{S}{ii}]~$\lambda6716,6731$). Spatial elements that did not satisfy this criterion in any of the required lines were excluded from the analysis.

To ensure that the emission originates from gas photoionized by young stars, we applied Baldwin-Phillips-Terlevich \citep[BPT;][]{Baldwin_1981} diagnostic diagrams using [\ion{O}{iii}]~$\lambda5007$ / H$\beta$ versus [\ion{N}{ii}]~$\lambda6583$ / H$\alpha$. Pixels located below the \cite{Kauffmann_2003} demarcation curve were classified as star-forming and were retained for further analysis. This selection reduces contamination from ionization sources other than massive stars (e.g., AGNs, HOLMES, or shock excitation), for which standard calibrations are not applicable.

Gas-phase oxygen abundances were computed using the strong-line calibration of  \cite{Pilyugin_2016}, based on nitrogen, sulfur, and oxygen emission lines relative to \(\rm H_{\beta}\).

The following line ratios were used:
\begin{align*}
   N_2 &= \frac{I_{[\ion{N}{ii}]\lambda6548+\lambda6584}}{I_{\rm H\beta}},\\
   S_2 &= \frac{I_{[\ion{S}{ii}]\lambda6717+\lambda6731}}{I_{\rm H\beta}},\\
   R_3 &= \frac{I_{[\ion{O}{iii}]\lambda4959+\lambda5007}}{I_{\rm H\beta}}
\end{align*}
The calibration adopts a two-branch formulation, with the lower and upper metallicity branches separated by $\rm \log N_2 = -0.6$. For pixels on the lower branch ($\rm \log N_2 \leq -0.6$), the oxygen abundance is given by:
\begin{align*}
\rm 12 + \log(O/H)_L &= 8.072 + 0.789\log(R_3/S_2) + 0.726 \log N_2 \\
& + \left( 1.069 - 0.170 \log (R_3/S_2) + 0.022 \log N_2 \right) \\
& \times \log S_2  ,
\end{align*}
while for the upper branch ($\log N_2 > -0.6$), it is given by:
\begin{align*}
   \rm 12 + \log(O/H)_U &= 8.424 + 0.030\log(R_3/S_2) + 0.751 \log N_2 \\
& + \left( -0.349 + 0.182 \log (R_3/S_2) + 0.508 \log N_2 \right) \\
& \times \log S_2 .
\end{align*}

This calibration has been specifically optimized to minimize its dependence on the ionization parameter and gas pressure, which makes it particularly suitable for spatially resolved analyses. It has been shown to provide reliable abundances at sub-solar metallicities \citep[e.g.,][]{DeVis_2019}

In addition to the spatially resolved abundances, we derived the integrated (global) metallicity for each galaxy, see Table~\ref{tab:Integrated}. To ensure consistency with the resolved analysis, only pixels classified as star-forming in the BPT and those located within the optical radius $R_{25}$ were considered. 
The emission line fluxes from these selected pixels were summed to obtain integrated line fluxes, and the same \cite{Pilyugin_2016} calibration was applied to the resulting line ratios.

In this work, we adopt the solar oxygen abundance $12 + \log(\text{O/H})_{\odot} = 8.69$ from \citet{Asplund_2009}, and the relative metallicity is defined as $Z/Z_{\odot} = 10^{(12 + \log(\text{O/H})) - 8.69}$.

\subsection{Atomic, molecular, and total gas}
\label{sec:method_gas}

Atomic ($\Sigma_{\mathrm{HI}}$) and molecular gas ($\Sigma_{\mathrm{H_2}}$) surface density maps were derived from the \text{HI} and CO data,
respectively. \(\mathrm{HI}\) intensity maps were converted to brightness temperature units following the standard prescription \citep{Schruba_2011}. 
Assuming optically thin 21\,cm emission, $\Sigma_{\mathrm{HI}}$ was computed as:
\begin{equation}
\Sigma_{\mathrm{HI}} = 0.02 \; I_{21\,\mathrm{cm}},
\end{equation}
where $I_{21\,\mathrm{cm}}$ is in units of $\rm K\,km\,s^{-1}$ and $\Sigma_{\mathrm{HI}}$ is expressed in $\rm M_{\odot}\,pc^{-2}$. The derived \(\Sigma_{\mathrm{HI}}\) includes already the contribution of helium and gas metals.

$\Sigma_{\rm H_2}$ was derived from the $^{12}$CO$(2-1)$ or $^{12}$CO$(1-0)$ integrated intensity maps, assuming optically thick CO emission. 
$\Sigma_{\rm H_2}$ is related to the $J=1 \rightarrow 0$ line intensity ($I_{\rm CO(1-0)}$) as:
\begin{equation}
\Sigma_{\rm H_2} = \alpha_{\rm CO} \, I_{\rm CO(1-0)},
\label{eq:sigma_h2}
\end{equation}
where $\alpha_{\rm CO}$ is the CO-to-$\rm H_2$ conversion factor in units of $\rm M_{\odot}\,pc^{-2}\,(K\,km\,s^{-1})^{-1}$. 
This formulation is equivalent to the global relation $M_{\rm {H_2}} = \alpha_{\rm CO} L'_{\rm CO}$ \citep{Bolatto_2013}, as the surface density represents the mass per unit area and the intensity $I_{\rm CO}$ represents the luminosity per unit area\footnote{The CO-to-$\rm H_2$ conversion factor is also defined as the ratio between the H$_2$ column density and the CO intensity, $X_{\rm CO} = N_{\rm H_2}/I_{\rm CO}$ [cm$^{-2}$ (K km s$^{-1}$)$^{-1}$]. 
The factors $\alpha_{\rm CO}$ and $X_{\rm CO}$ are related by $X_{\rm CO} \approx 6.3 \times 10^{19} \, \alpha_{\rm CO}$.}. 
For our analysis based on $^{12}$CO$(2-1)$ observations, we converted the observed intensities to the $J=1 \rightarrow 0$ scale adopting a constant line ratio $R_{21} = I_{\rm CO(2-1)}/I_{\rm CO(1-0)} = 0.7$ \citep[e.g.,][]{Schruba_2011, Leroy_2013, Casasola_2015, Yajima_2021, Pas_2025}.

Since $\alpha_{\rm CO}$ is not universal, we consider three different prescriptions for it (see Sect.~\ref{sec:alpha_CO}). 
This choice is critical for DGR and DMR studies, as variations in $\alpha_{\rm CO}$ (especially in low-metallicity regimes) directly impact the inferred molecular mass content. 
To ensure consistency with the atomic gas component, $\Sigma_{\rm H_2}$ values include the contribution of helium and gas metals.
This is achieved by adopting an $\alpha_{\rm CO}$ that already incorporates a factor of 1.36.

The total gas surface density, $\Sigma_{\rm gas}$, was computed as the sum of $\Sigma_{\rm HI}$ and $\Sigma_{\rm H_2}$:
\begin{equation}
    \Sigma_{\rm gas} = \Sigma_{\rm HI} + \Sigma_{\rm H_2}.
\end{equation}

\subsubsection{The CO-to-H$_2$ conversion factor}
\label{sec:alpha_CO}

The non-universality of $\alpha_{\rm CO}$ arises from its strong dependence on the physical conditions of the ISM, being primarily driven by variations in metallicity.
The origin of this dependence lies in the different shielding requirements of gas species.

While molecular hydrogen can efficiently self-shield against dissociating radiation once a sufficient column density is reached, the CO molecule is more vulnerable to photodissociation and relies heavily on dust shielding to remain abundant.
Consequently, in low-metallicity environments where the dust content is reduced, a significant fraction of $\rm H_2$ may reside in "CO-dark" regions.
In such conditions, a given CO luminosity corresponds to a larger molecular gas mass, resulting in higher \(\alpha_{\rm CO}\) values \citep[e.g.,][]{Wolfire_2010,Glover_2011}.
This effect becomes more pronounced below the metallicities of \(\sim 1/3 - 1/2\;Z_{\odot}\), where shielding strongly limits the survival of CO \citep[e.g.,][]{Leroy_2011b,Bolatto_2013,Hu_2021}.

Additionally, variations in gas surface density and radiation field strength, gas kinematics, and optical depth may further contribute to deviations of \(\alpha_{\rm CO}\) from a constant value representative of solar-metallicity environments. For example, \citet{Teng_2022} found substantial spatial variations in \(\alpha_{\rm CO}\) across NGC~3351, with significantly lower values in the inflow arms than in the central regions. They attributed these variations to changes in velocity dispersion and optical depth of CO. Similarly, \citet{Yasuda_2023} reported systematic radial variations in \(\alpha_{\rm CO}\) across nearby spiral galaxies, with lower values in inner high-SFR regions and enhanced values (larger by factors of 1.3 to 5.3) in the outer disks.
 
The exact functional form of the metallicity dependence of \(\alpha_{\rm CO}\) remains uncertain, and a variety of empirical calibrations have been proposed in the literature \citep[e.g.,][]{Genzel_2012,Schruba_2012,Sandstrom_2013,Hunt_2015,Amorin_2016,Accurso_2017, Bisbas_2025}. 
The reported slopes of the \(\alpha_{\rm CO}\)--Z relation vary significantly, which directly affects the inferred molecular gas masses.

To evaluate how the choice of \(\alpha_{\rm CO}\) influences our derived gas mass estimates and the resulting DGR and DMR relations, we adopt three commonly used prescriptions:

\noindent (i) \citet[][B13]{Bolatto_2013} -- As a reference case, we adopt a constant conversion factor representative of the MW:
\begin{equation}
    \alpha_{\rm CO, B13} = 4.35 \; \rm M_{\odot} \, pc^{-2} (K \, km \, s^{-1})^{-1}.
\end{equation}
This value is widely adopted for nearby spiral galaxies with approximately solar metallicity, corresponds to a standard Galactic $X_{\rm CO} = 2.0 \times 10^{20} \; \rm cm^{-2} (K \, km \, s^{-1})^{-1}$ and explicitly includes a factor of 1.36 to account for helium and heavier elements.

\noindent (ii) \citet[][A16]{Amorin_2016} -- To account for the reduced
CO emissivity expected at sub-solar metallicity, we adopt the
$Z$-dependent scaling propose by \cite{Amorin_2016}, 
derived from CO observations of  21 blue compact dwarf galaxies (BCDs), spanning a wide range of physical properties, with $12 + \log(\text{O/H}) \sim 7.69 - 8.86$, $\log(\Sigma_{\rm SFR} / [\rm M_{\odot} \, yr^{-1} \, kpc^{-2}]) \sim -1.77$ to $-0.41$, and $M_* \sim 0.07 - 15.85 \times 10^9 \,\rm M_{\odot}$.
By examining the correlation between the molecular gas depletion time (defined as the timescale required to exhaust the current molecular reservoir at the observed SFR, $\tau_{\rm H_2} = M_{\rm H_2}/\rm{SFR}$), 
the metallicity, and the specific SFR (defined as the SFR per unit of stellar mass, $\rm{sSFR} = \rm{SFR}/M_{\star}$), they infer a $Z$-dependent $\alpha_{\rm CO}$ given by:
\begin{equation}
    \alpha_{\rm CO, A16} = 5.0 \times \left(\frac{Z}{Z_{\odot}}\right)^{-1.5} \; \rm M_{\odot} \, pc^{-2} (K \, km \, s^{-1})^{-1}.
\end{equation}
Compared to the constant Galactic value adopted in B13, this scaling predicts an increase of \(\alpha_{\rm CO}\) towards lower metallicity.

\noindent (iii) \citet[][S12]{Schruba_2012} -- As an alternative scaling, we adopt the empirical calibration proposed by \cite{Schruba_2012}. This relation is based on stacked CO measurements of nearby low-mass, low luminosity galaxies (\(L'_{\rm CO (2-1)} \leq 28 \times 10^6 \; \rm Kkm^{-1}pc^2\)). 
By comparing CO luminosity to SFR and assuming an approximately constant $\tau_{\rm H_2}$,
they showed that galaxies with \(Z \sim 1/2 - 1/10\; Z_{\odot}\) are significantly fainter in CO relative to their SFRs \citep[e.g.,][]{Bigiel_2008}. Interpreting this deficit as evidence for CO becoming increasingly poor tracer of H$_2$ at low $Z$, they derived a steep metallicity dependence:
\begin{equation}
    \alpha_{\rm CO, S12} = 8.0 \times \left(\frac{Z}{Z_{\odot}}\right)^{-2} \; \rm M_{\odot} \, pc^{-2} (K \, km \, s^{-1})^{-1}.
\end{equation}

\noindent Relative to the A16 prescription, the S12 formulation exhibits a stronger dependence on metallicity and thus predicts a more rapid increase of \(\alpha_{\rm CO}\) towards lower metallicity, reflecting an increasing contribution from CO-faint molecular gas.
Using these two latter formulations allows us to evaluate the impact of the assumed metallicity dependence on our derived DGR--$Z$ and DMR--$Z$ trends. By adopting both A16 and S12, we can account for the uncertainties related to their different assumptions regarding $\tau_{\rm H_2}$ in metal-poor environments.

\subsection{Dust-to-gas and dust-to-metal ratios}
\label{sec:ratios}

The DGR is defined as the ratio between $\Sigma_{\rm dust}$ and $\Sigma_{\rm gas}$:
\begin{equation}
    \rm DGR = \frac{\Sigma_{\rm dust}}{\Sigma_{\rm gas}}.
\end{equation}
This quantity provides a measure of the relative abundance of dust grains within the ISM.

To estimate the DMR\footnote{Different definitions of the DMR in the literature primarily arise from how the $\Sigma_{\rm metals}$ is estimated.
Some studies approximate the total metal content using only gas-phase metallicity measurements from H\,\textsc{ii} regions \citep{Chiang_2021}.
Others define the metal budget as the sum of gas-phase and dust-phase metals by combining metallicity with dust measurements \citep[e.g.,][]{DeVis_2019,DeLooze_2020}.} 
we adopt the formulation of \cite{DeVis_2019} which explicitly accounts for metals locked in dust, providing a more complete census of the metal budget when combining dust and gas-phase measurements on resolved scales, thus we get:

\begin{equation}
    \rm DMR = \frac{\Sigma_{\rm dust}}{\Sigma_{\rm metals}},
    \label{eq:dmr}
\end{equation}
\noindent where the total metal mass surface density (\(\Sigma_{\rm metals}\)) is given by:
\begin{equation}
    \Sigma_{\rm metals} = f_Z \Sigma_{\rm gas} + \Sigma_{\rm dust},
    \label{eq:metal_fn}
\end{equation}
and $f_Z$ represents the metal mass fraction in the gas phase, defined as $f_Z = 27.36 \times 10^{(12 + \log(\text{O/H})) - 12}$. 
This is related to the relative metallicity $Z/Z_{\odot}$ defined in Sect.~\ref{sec:ratios} as $f_Z = (Z/Z_{\odot}) \times Z_{\odot, \text{mass}}$, where $Z_{\odot, \text{mass}}$ is the solar oxygen mass fraction (e.g., $0.0134$ for the \citeauthor{Asplund_2009} \citeyear{Asplund_2009} scale).
In this framework, the total metal budget includes both gas-phase metals and those locked in dust grains.

Both DGR and DMR were computed on a pixel-by-pixel basis, within regions where all four quantities ($\Sigma_{\rm HI}$, $\Sigma_{\rm H_2}$, $\Sigma_{\rm dust}$, $12+\log(\text{O/H})$) are detected with a S/N~$> 3$. 
For the gas-phase metallicity specifically, we required a S/N $> 3$ in all emission lines involved in the selected calibration, and ensuring that the pixel is classified as star-forming according to the BPT diagram.

This approach ensures a high and homogeneous statistical significance across our entire resolved dataset, preventing our physical correlations from being biased by noise-dominated pixels in any individual map. 
Consequently, non-detections and marginal signals are not treated as upper limits but are excluded from the analysis. Although this selection could introduce a selection bias against dust-poor or gas-poor regions, its impact is mitigated by our target selection (spiral galaxies with near-solar to solar metallicities, $12 + \log(\text{O/H}) \gtrsim 8.2$) and spectroscopic requirements. 
The criterion of a $S/N > 3$ in all emission lines needed for the metallicity calibration, combined with the strict BPT classification, represents the main factor limiting our final pixel statistics (as seen in the final $N_{\rm pix}$ values in Table~1). 
Furthermore, because the signal in all our maps ($\Sigma_{\text{H}_2}$, $\Sigma_{\text{dust}}$, and emission lines) decreases towards the outskirts of the galaxies, our joint $3\sigma$ threshold acts as a physical boundary. 
As a result, our analysis is restricted to the stellar-dominated optical disks ($R/R_{25} \le 1$), and for many galaxies it stops well before the optical boundary where the emission drops to the noise level (see radial profiles in Appendices \ref{appendix: radial prof dgr} and \ref{appendix: radial prof dmr}). 
Since we are not probing the diffuse, extremely low-metallicity outskirts of the galaxies, the potential selection bias of omitting upper limits is very limited, and our sample remains representative of the active star-forming ISM.

As an example, the resulting matched-resolution and regridded maps for NGC~5457 are presented in Appendix~\ref{appendix:ngc5457}, where the spatial distributions of $\Sigma_{\rm dust}$, $\Sigma_{\rm HI}$, $\Sigma_{\rm H_2}$, $\Sigma_{\rm M_*}$, DGR, and DMR are shown.
The gas-phase metallicity maps, with $\Sigma_{\rm SFR}$ contours, are presented for the all sample galaxies in Fig.~\ref{fig:metallicity_coverage}.

\section{Results}
\label{sec:results}
In this section, we present the resolved DGR--$Z$ and DMR--$Z$ scaling relations derived for our galaxy sample. 
We evaluate how these relations are affected by the choice of $\alpha_{\rm CO}$.
Furthermore, we examine the $L_{\rm CO(1-0)}$--SFR relation, taking into account the gas-phase metallicity and local ISM properties, to assess the reliability of CO as a molecular gas tracer across different galactic environments.
Our results are also compared with previous observational studies. 

\subsection{The resolved DGR--$Z$ relation}
\label{sec:result_dgr}

\begin{figure*}[htbp]
    \centering
    \includegraphics[width=1\linewidth]{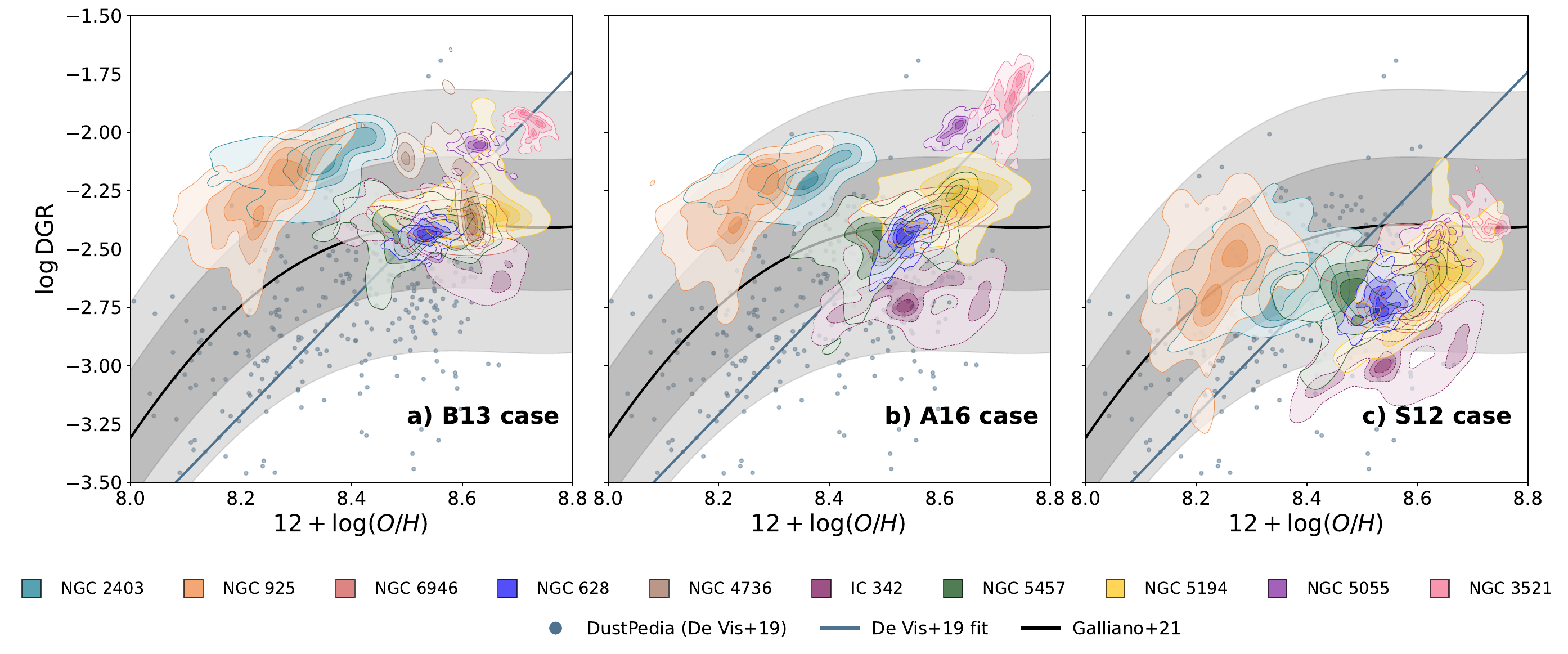}
    \caption{
    Resolved DGR as a function of gas-phase metallicity for our galaxy sample, computed using three different \(\alpha_{\rm CO}\) prescriptions: a) B13 (MW-type), b) A16 (compact dwarfs-type), c) S12 (low mass, low CO luminosity). Colored contours show the density distribution for each galaxy in the resolved measurements; galaxy colors are defined in the legend, ordered by increasing stellar mass.
    Gray points represent the DustPedia sample, and the gray line indicates the corresponding fit to these points from \cite{DeVis_2019}, while the black curve represents the relation from \cite{Galliano_2021}, with shaded region showing \(1\sigma\) and \(2\sigma\) scatter in the relation. 
    NGC~4736 is shown only in panel a) due to the limited spatial coverage of its metallicity map (see Fig.~\ref{fig:metallicity_coverage}). IC~342 is shown with dashed contours to highlight its starburst nature and extreme ISM conditions, as discussed later (see Sect.~\ref{sec:co-sfr}).}
    \label{fig:dgr-Z}
\end{figure*}

Figure~\ref{fig:dgr-Z} shows the DGR as a function of metallicity, 
comparing the results obtained using the three $\alpha_{\rm CO}$ prescriptions: B13 (left panel), A16 (middle panel), and S12 (right panel). The colored contours trace the density distribution of the resolved measurements for each galaxy, representing variations across the galactic disk.
The metallicity covers a relatively limited range of $12 + \log(\text{O/H}) \sim 8.1$ --  $8.8$, corresponding to sub-solar up to nearly solar abundances, as commonly found in nearby spiral galaxies \citep[e.g.,][]{Bresolin_2009,Stanghellini_2014,Stanghellini_2015,Casasola_2022,Lara-Lopez_2023,Pilyugin_2025,Vale_25}. 
Within this interval, \( \log(\mathrm{DGR}) \) spans roughly 1 dex, from approximately \(-3\) to \(-2\). Overall, for the regions we probe and regardless of the assumed $\alpha_{\rm CO}$ prescription, the resolved DGR–$Z$ relation exhibits some common features: a relatively shallow trend with metallicity and a considerable galaxy-to-galaxy scatter at fixed metallicity. 

In Fig.~\ref{fig:dgr-Z}, we compare our resolved DGR--$Z$ distribution with the global relations from the literature. 
We show the individual galaxy data from the DustPedia sample \citep[][gray dots and gray curve]{DeVis_2019} alongside the relations from \citet[solid black line]{Galliano_2021}. The \citet{DeVis_2019} relation is an empirical fit to global DustPedia galaxy measurements, while the \citet{Galliano_2021} relation is based on an analytical dust evolution model.
Our measurements lie within the galaxy-to-galaxy scatter of the DustPedia points, though they exhibit a significantly tighter distribution. 
However, regardless of the individual $\alpha_{\rm CO}$ prescription adopted, our data points tend to trace shallower trends than that derived by \citet{DeVis_2019}. 

A more detailed inspection of Fig.~\ref{fig:dgr-Z} reveals that the DGR–$Z$ relation is largely  dependent on the assumed $\alpha_{\rm CO}$ prescription. 
In the left panel, when using the constant  $\alpha_{\rm CO}$ from B13 (MW-type), the resolved measurements generally fall within the ranges reported in previous studies; however, pronounced galaxy-to-galaxy variations are present. 
Systems such as NGC~5055 and NGC~3521 lie towards higher DGR values ($\log(\text{DGR}) \sim -2.0$) relative to the main distribution. 
The lower-metallicity ($12 + \log(\text{O/H}) < 8.3$) galaxies in our sample (NGC~2403 and NGC~925) also show relatively elevated DGR values ($\log(\text{DGR}) \sim -2.2$) for their metallicities relative to the bulk of the sample.

In the middle panel of Fig.~\ref{fig:dgr-Z}, when adopting the metallicity-dependent $\alpha_{\rm CO}$ from A16,
the resolved measurements show a similar overall distribution; however, systematic shifts are observed in specific regions of the DGR--$Z$ plot. 
Systems such as NGC~5055 and NGC~3521 exhibit even more discrepant DGR values (up to $\log(\text{DGR}) \sim -1.75$) compared to the average trend. 
The lower-metallicity galaxies (NGC~2403 and NGC~925) maintain relatively elevated DGR values ($\log(\text{DGR}) \sim -2.2$), showing that the A16 prescription does not significantly reduce the scatter for these specific objects. 
The impact of this prescription is most evident in the inner regions of some galaxies (NGC~5055, NGC~5457, NGC~6946, IC~342), where the DGR shifts toward higher values as shown in the corresponding radial profiles (see Appendix~\ref{appendix: radial prof dgr}).
This occurs because these galaxies reach higher metallicities in their inner regions, where $\alpha_{\rm CO}$ is expected to decrease according to the A16 prescription. 
Consequently, allowing for a metallicity-dependent $\alpha_{\rm CO}$ modifies the inferred molecular gas masses, producing systematic shifts in the derived DGR, particularly in metal-rich inner regions.

In the right panel of Fig.~\ref{fig:dgr-Z}, when adopting the S12 \(\alpha_{\rm CO}\), the resolved measurements lead to a reduced scatter at a fixed metallicity compared to the B13 and A16 prescriptions, while showing a systematic shift toward lower DGR values across the full metallicity range.
This is due to the fact that the S12 prescription, in addition to a strong dependence on metallicity, adopts a comparatively high normalization, resulting in systematically higher \(\alpha_{\rm CO}\) values. 
Consequently, the inferred molecular gas surface densities are significantly enhanced, leading to the observed decrease in DGR.
This effect is most pronounced at sub-solar metallicities, where the increased molecular gas mass causes the low-CO-luminosity systems (specifically NGC~925 and NGC~2403) to fall more in line with the rest of the sample. 
As a result, the S12 prescription yields a noticeably flatter DGR--$Z$ relation and a narrower overall spread compared to the B13 and A16 prescriptions.
Taken together, these results indicate that while metallicity is a first-order driver of the DGR, systematic uncertainties in the gas mass estimates -- specifically the choice of $\alpha_{\rm CO}$ -- play a critical role in shaping the observed slope and scatter of the relation.

\subsection{The resolved DMR--$Z$ relation}
\begin{figure*}[htbp]
    \centering
    \includegraphics[width=1\linewidth]{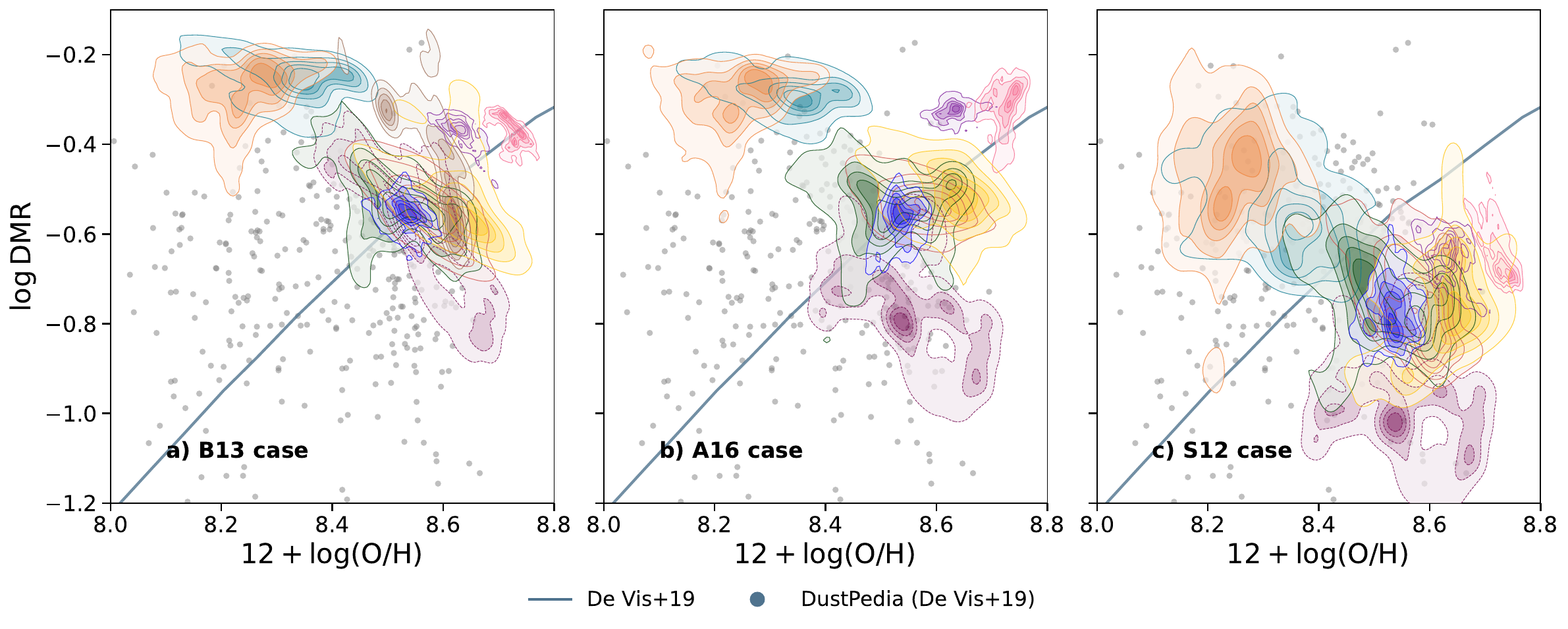}
    \caption{Resolved DMR as a function of gas-phase metallicity for our galaxy sample with three \(\alpha_{\rm CO}\) prescriptions as in Fig.~\ref{fig:dgr-Z}. Color of density distribution of each galaxy is same as in Fig~\ref{fig:dgr-Z}. The gray curve shows the DMR-metallicity trend found in \cite{DeVis_2019}. Gray points represent the DustPedia sample from \cite{DeVis_2019}. 
    }
    \label{fig:dmr-Z}
\end{figure*}

Figure~\ref{fig:dmr-Z} shows the DMR as a function of metallicity, for the three $\alpha_{\rm CO}$ prescriptions.
Within the metallicity range of our sample, $\log(\text{DMR})$ typically varies between $\sim -1$ and $-0.2$.
Across all panels, DMR shows a (weak) downward trend with increasing metallicity, with substantial galaxy-to-galaxy variations at fixed metallicity.

We additionally overlay the relation from \citet{DeVis_2019} which shows an overall increasing trend of DMR with metallicity. However, this increase is driven by the inclusion of low-metallicity, less evolved systems ($12+\log(\rm O/H) \lesssim 8.2$) in their sample (see their Fig.~10), a regime not probed by our data. In the higher-metallicity range relevant to our sample, \citet{DeVis_2019} instead report an approximately constant DMR, with $\log(\rm DMR) \sim -0.67$, in closer agreement with our measurements.

As for the DGR--$Z$ relation, a closer look at Fig.~\ref{fig:dmr-Z} shows that the DMR--$Z$ distribution depends on the assumed $\alpha_{\rm CO}$ prescription.
In the left panel, using the B13 \(\alpha_{\rm CO}\) prescription, the distribution exhibits a broad range of $\log(\text{DMR})$ values at fixed metallicity, but the galaxy-averaged values suggest two broad regimes. A subset of galaxies (NGC~2403, NGC~925, NGC~5055, NGC~3521) occupies systematically higher DMR values, with mean $\log(\text{DMR}) \approx -0.37$ to $-0.27$, corresponding to approximately 55\% - 45\% metals being locked in dust grains. In contrast, the remaining systems cluster around lower values, with mean $\log(\text{DMR}) \approx -0.53$, which corresponds to 30\% metals being locked in dust. Many individual galaxies exhibit negative internal gradients (IC~342, NGC~5194, NGC~6946, NGC~5457, NGC~3521). The inner, high-metallicity regions, tend to show slightly lower DMR compared to their outer parts. However, these trends are not uniform across the sample. This behavior should be interpreted with caution because DMR and metallicity are not fully independent quantities (see Eq.~\ref{eq:dmr} and \ref{eq:metal_fn}). Therefore, the observed trend cannot be interpreted solely as a direct decrease in the fraction of metals locked into dust with increasing metallicity. \cite{Casasola_2022}, found for their resolved analysis using metallicities derived from radial averages, that the DMR--$Z$ relation under constant \(\alpha_{\rm CO}\) prescription is flat or weakly positive (see their Fig. 11). This difference compared to our results likely reflects the additional information provided by resolved 2D metallicity maps from IFU spectroscopy, which capture local ISM variations and chemical inhomogeneities that are averaged out when using radial gradients.

In the middle panel of Fig.~\ref{fig:dmr-Z}, when adopting the metallicity-dependent A16 \(\alpha_{\rm CO}\) prescription, the overall DMR--$Z$ distribution remains broadly similar to the B13 case, but the separation between two regimes is more clearly defined. The upper group (NGC~2403, NGC~925, NGC~5055, and NGC~3521) shows elevated DMR values and is centered around $\log(\text{DMR}) \approx -0.3$, while the rest of the sample clusters tightly near $\log(\text{DMR}) \approx -0.55$. The lowest values are found for IC~342, with a mean value of $\log(\text{DMR}) \approx -0.7$, and its inner regions exhibiting especially low DMR. Compared to B13 case, a notable difference is the flattening of internal gradients within galaxies, which is consistent with a scenario in which the fraction of metals locked into dust approches a near-constant value in the metal-rich ISM, reducing the sensitivity of DMR to local variations in metallicity. 

In the right panel of Fig.~\ref{fig:dmr-Z}, when adopting the S12 prescription, we observe a noticeable departure of the DMR--$Z$ relation compared to previous cases. The distribution is shifted towards lower values across the full sample, with galaxy-averaged means spanning $\log(\text{DMR}) \approx -1.0$ to $-0.5$, with a concentration of galaxies towards $\log(\text{DMR}) \approx -0.75$. IC~342 defines the lower end of the distribution (mean $\log(\text{DMR}) \approx -0.1 $), including very low values in its inner regions. The separation into two regimes seen in previous cases is no longer apparent; instead the sample follows a pronounced negative trend with metallicity, which is also reflected within individual galaxies: the inner, more metal-rich regions exhibit lower DMR values compared to the outer parts, leading to an overall steepening of the DMR--$Z$ relation.
This behavior is likely driven by the adopted \(\alpha_{\rm CO}\) prescription, which introduces a stronger metallicity dependence in the inferred molecular gas masses. 

In summary, our results demonstrate that both the resolved DGR--$Z$ and DMR--$Z$ relations are sensitive to the choice of \(\alpha_{\rm CO}\) prescription. The constant B13 prescription produces declining DMR--$Z$ trends in several galaxies, this behavior should be interpreted considering the mathematical coupling between DMR and metallicity. The metallicity-dependent prescriptions further modify this relation, but the impact depends on the adopted scaling. The A16 prescription reduces the internal DMR gradients compared to B13, resulting in a flatter trend consistent with an approximately constant fraction of metals locked into dust in the metal-rich regime. In contrast, the stronger metallicity dependence of the S12 prescription introduces a more pronounced negative trend, both globally and within individual galaxies, by increasing the inferred molecular gas masses in lower-metallicity regions. 
These results highlight that uncertainties in the molecular gas conversion factor can significantly affect the inferred scaling relations stressing the need to adopt a hybrid \(\alpha_{\rm CO}\) approach tailored to the different galactic environments within our sample.

 \begin{figure}
    \centering
    \includegraphics[width=1\linewidth]{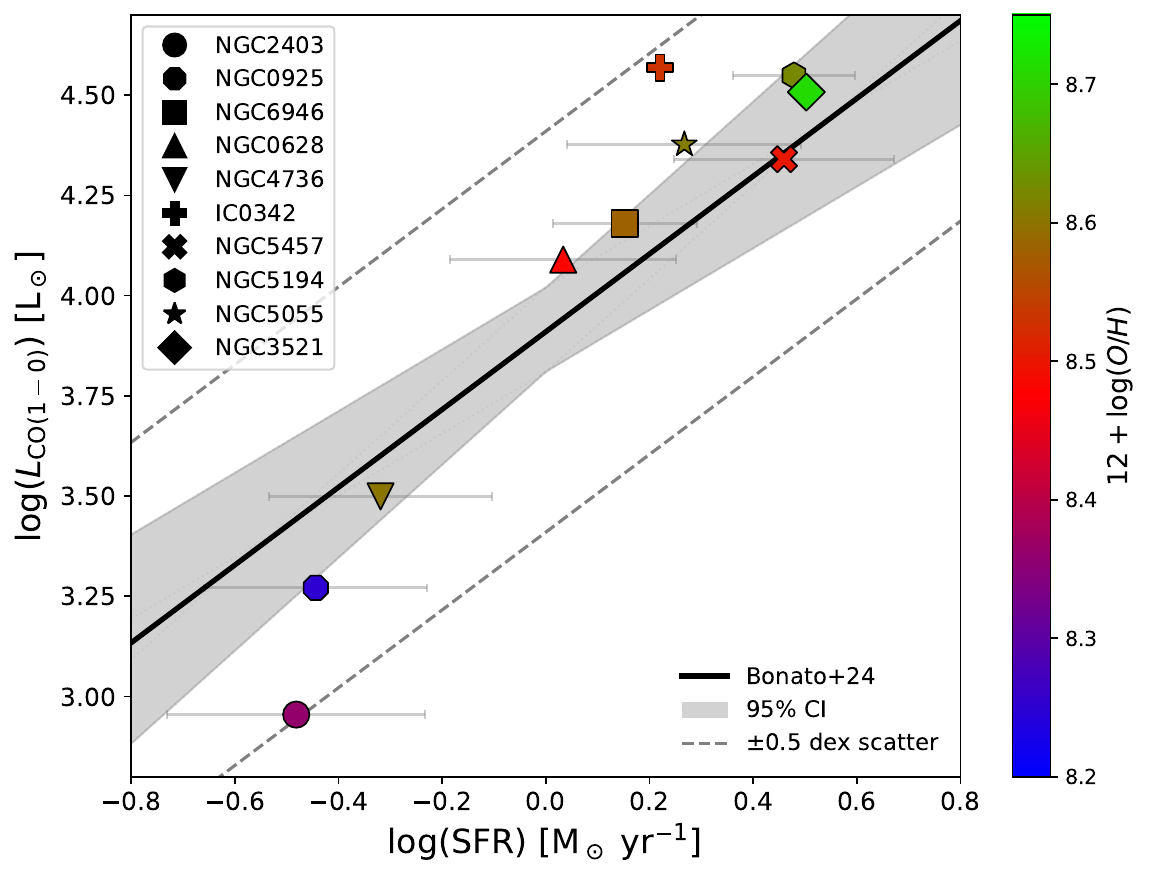}
    \caption{CO luminosity as a function of star formation rate for the galaxy sample. The solid black line represents the empirical relation from \cite{Bonato_2024}, with the gray shaded area representing the robust 95$\%$ confidence intervals of the fit. The dashed lines indicate the $\pm0.5$ dex intrinsic scatter of the reference galaxy population from \cite{Bonato_2024}.
    }
    \label{fig:lco_sfr}
\end{figure}

\subsection{CO luminosity--SFR relation and implications for $\alpha_{\rm CO}$}
\label{sec:co-sfr}
To investigate the physical feasibility of using a hybrid approach for $\alpha_{\rm CO}$, we examine the relation between $L_{\rm CO(1-0)}$ and SFR as a diagnostic of how effectively CO traces the molecular gas reservoir across our sample. 
Under the assumption of an approximately constant $\tau_{\rm H_2}$, star-forming galaxies are expected to follow a near-linear $L_{\rm CO(1-0)}$--SFR relation \citep[e.g.,][]{Bigiel_2008,Leroy_2013}.
Deviations from this linearity can therefore provide insight into variations in CO emissivity,\textbf{ $\tau_{\rm H_2}$}, and the suitability of the adopted \(\alpha_{\rm CO}\).
In Fig.~\ref{fig:lco_sfr}, we show $L_{\rm CO(1-0)}$ as a function of SFR for our galaxies, together with the empirical relation derived from a large subset (388 galaxies) of the DustPedia sample by \cite{Bonato_2024}.
The figure also displays the intrinsic scatter of the \cite{Bonato_2024} sample (\(\sim 0.5\) dex, dashed gray lines) and the 95\% confidence interval (CI, gray shaded region) associated with the best-fit slope and intercept. The intrinsic dispersion of $\sim0.5$ dex of the reference sample demonstrates that a significant galaxy-to-galaxy variation exists around the average relation over the entire SFR range probed. Therefore, the position of an individual galaxy on this diagram cannot be interpreted independently, but must be considered together with its metallicity and ISM conditions. In this context, the $L_{\rm CO(1-0)}$–SFR plane provides a useful diagnostic to identify systems with potentially different \(\alpha_{\rm CO}\), while the physical properties of each galaxy determine the origin of the observed offset.

Most galaxies in our sample are consistent with the \citet{Bonato_2024} relation within the observed scatter, following a linear trend consistent with a typical $L_{\rm CO(1-0)}$ to SFR ratio of $\rm \sim 8.1 \times 10^3 \, L_{\odot} (M_{\odot} \, {\rm yr}^{-1})^{-1}$, indicating that their CO luminosities are broadly consistent with those of the general star-forming population.
However, IC~342 and NGC~2403, are separated from the 95\% CI, occupying opposite regimes of the relation. IC~342 lies above the relation, exhibiting enhanced CO emission relative to its SFR, while NGC~2403 is located well below the relation, showing a lower $L_{\rm CO}$ than expected for its SFR.

The offset of IC~342 by a factor of \(\sim 2.8\) from the relation is consistent with a longer $\tau_{\rm H_2}$ or, alternatively, with enhanced CO emissivity. 
IC~342 is known to host a dense and prominent central starburst region, where high gas densities, elevated kinetic temperatures, and increased velocity dispersion can influence the molecular gas conditions \citep[see e.g.,][]{Meier_2011, Pan_2014,I_2020}. These physical conditions enhance CO excitation, boosting $L_{\rm CO}$ relative to SFR \citep[e.g.,][]{Querejeta_2023}.
The latter interpretation is supported by \citet{Tailor_2025}, who found that the dust heating across the disk of IC~342 is mainly driven by evolved stellar populations rather than a young UV radiation field.
The absence of an intense UV field outside the starburst core prevents the large-scale photodissociation of CO molecules, confirming that there is no physical basis to expect a severe CO-dark gas fraction in this system.
This implies that variations in its $\alpha_{\rm CO}$ factor are likely regulated by local gas density and dynamical state rather than purely chemical effects.
Consistently, \citet{Chiang_2021} showed that an $\alpha_{\rm CO}$ calibration based solely on metallicity cannot fully capture the complex ISM conditions of IC~342, and that incorporating the total surface density offers a more accurate description. However, deriving such galaxy-specific, density-dependent corrections is beyond the scope of the present work. Therefore, we adopt the standard metallicity-dependent A16 prescription for IC~342, treating it as a reasonable approximation despite the central starburst.

Several other galaxies in our sample (NGC~925, NGC~5055, NGC~628, NGC~6946, and NGC~5194) are located near the edge of the 95\% CI. All of these except NGC~925, exhibit positive offsets from the $L_{\rm CO}$–SFR relation by a factor of $\lesssim 1.6$. These galaxies do not show extreme conditions like the central starburst of IC~342. Their mild offsets are likely associated with small variations in molecular gas properties, CO excitation, $\tau_{\rm H_2}$, or SFR uncertainties. Their near-solar metallicities and standard ISM conditions do not motivate a departure from the A16 prescription that we adopt for this group of galaxies.

Conversely, NGC~2403 and NGC~925 represent a chemically distinct regime within our sample, characterized by sub-solar metallicities ($12 + \log(\rm O/H) = 8.36$ and $8.25$, respectively). 
In this regime, the A16 prescription may not be adequate to account for a significant CO-dark molecular gas component, motivating us to adopt the steeper metallicity-dependent S12 prescription. 
NGC~2403 exhibits a CO deficit, with a lower $L_{\rm CO}$ by a factor of $\sim 3.1$ compared with the best-fit relation. Such negative offsets can be interpreted in two ways.  First, they may reflect enhanced star formation efficiency (SFE = $1/\tau_{\rm H_2}$, i.e. shorter depletion timescales), which are expected in lower-mass systems where galaxy downsizing can lead to more efficient star formation. In this context, \citet{Dib_2011} introduces a metallicity dependent feedback model in which the winds from massive OB stars are strongly metallicity dependent, with the wind power scaling approximately linearly with the metallicity. This results in longer gas expulsion timescales in low-metallicity star-forming regions and consequently higher SFEs. Second, the offset may indicate that CO traces only a fraction of the total molecular gas reservoir. In low-metallicity environments in the local Universe, the latter scenario is often favored.
In such low-metallicity environments, reduced dust shielding allows UV radiation to penetrate deeper into molecular clouds, leading to efficient CO photodissociation while $\rm H_2$ remains self-shielded \citep[e.g.,][]{Wolfire_2010, Schruba_2012, Bolatto_2013}. 
NGC~925 represents a similar case; its $L_{\rm CO}$ is lower than the relation by a factor of $\sim 1.6$, locating it near the lower boundary of the 95\% CI. 
This interpretation is supported by \citet{Tailor_2025}, who found that dust heating in both NGC~2403 and NGC~925 is mainly driven by young stellar populations. 
The resulting local UV radiation field, combined with the reduced dust shielding, significantly increases the photodissociation of CO, providing a solid physical basis for large CO-dark gas fractions in both systems.

We also note that NGC~2403 and NGC~925 differ in their environmental properties, which likely explains their different relative positions within the $L_{\rm CO}$–SFR plane despite their similar metallicity.
NGC~2403 is an interacting member of the M~81 group, and its more pronounced CO deficit is likely driven by the combination of its sub-solar metallicity and environmental effects that may have altered and reduced its molecular gas reservoir \citep[see e.g.,][]{Casasola_2004}. Conversely, NGC~925 is an isolated system, and its milder CO deficit is likely a direct manifestation of its low metallicity alone, causing it to remain closer to the 95\% CI boundary.

Further justification for adopting the S12 prescription for NGC~2403 and NGC~925 comes from the nature of the calibration itself. S12 was calibrated using low-mass, CO-faint dwarf galaxies in the Local Group (see Sect.~\ref{sec:alpha_CO}). Within our sample, NGC~2403 and NGC~925 are the only two systems approaching this low-$L_{\rm CO}$ regime, whereas the remaining galaxies display $L_{\rm CO}$ well above the baseline used for that calibration (Table~\ref{tab:Integrated}).
Additional independent support comes from \cite{Garduno2026}, who derived $\alpha_{\rm CO}$–$Z$ relations for a sample of Metal-THINGS galaxies. 
For NGC~2403 and NGC~925, they find a remarkably steep metallicity dependence with a slope of $\sim -5$. This implies that $\alpha_{\rm CO}$ increases much more rapidly toward lower metallicities than in standard prescriptions, consistent with the presence of a significant CO-dark molecular gas component in these systems.

\section{Discussion}
\label{sec:discussion}

\begin{figure}
    \centering
    \includegraphics[width=1\linewidth]{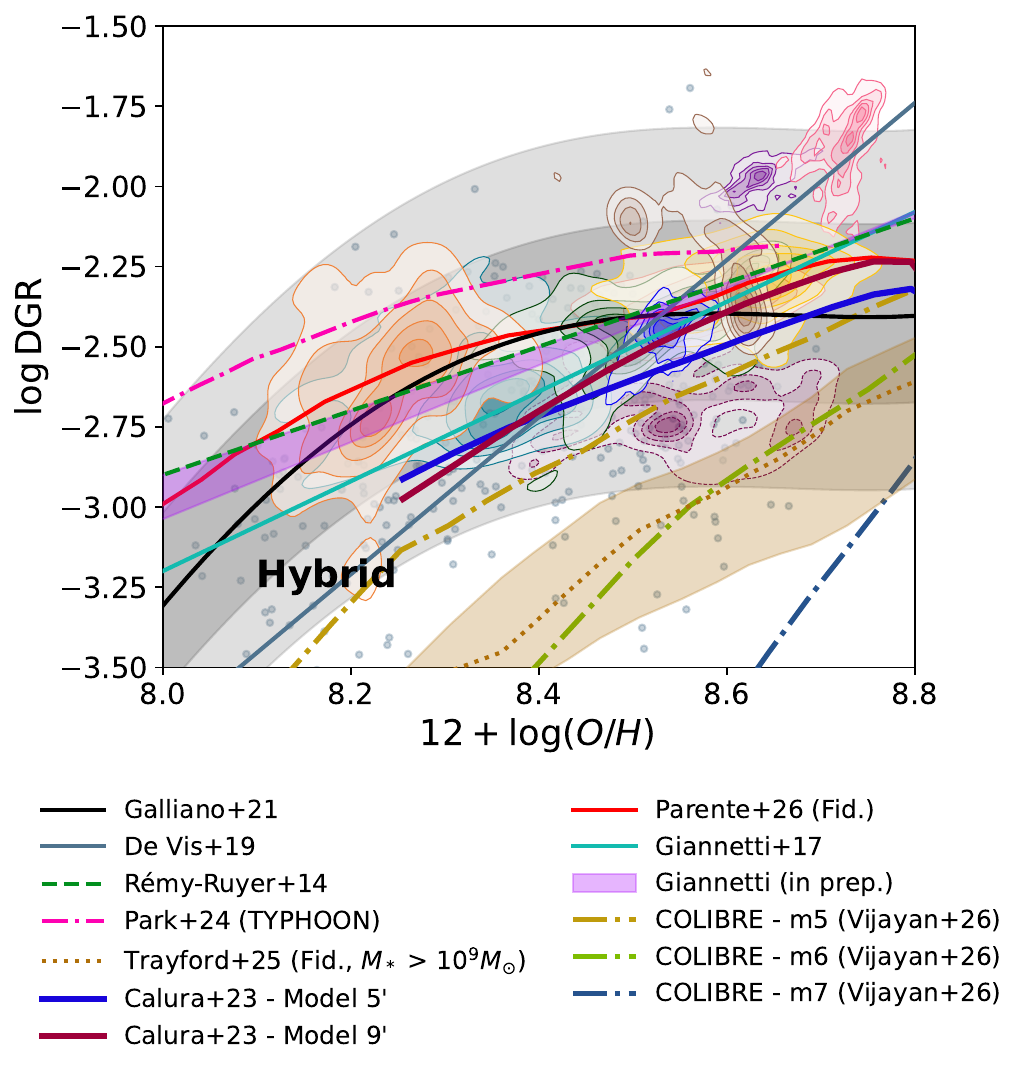}
    \caption{Resolved DGR as a function of metallicity for our galaxy sample using hybrid \(\alpha_{\rm CO}\) prescription (see Sect~\ref{sec:discussion}). Color of density distribution of each galaxy is same as in Fig.~\ref{fig:dgr-Z}, and the \cite{DeVis_2019} and \cite{Galliano_2021} relations are shown with gray and black curves, respectively, as in Fig.~\ref{fig:dgr-Z}. Overlaid curves illustrate a range of observational and theoretical predictions: the high-metallicity branch from \cite{RemyRuyer_2014} (green dashed), the resolved relation from \cite{Park_2024} (pink dash-dotted), MW relation from \cite{Giannetti_2017} (cyan solid), together with the MW $Z^{1-1.2}$ scaling range from Giannetti et al. (in prep., purple shaded region) and simulation-based relations from \cite{tray_2025} (brown dotted), and \cite{Parente_2026} (red solid). We also show the different prediction of COLIBRE simulation in dash-dot lines (yellow: m5 , greem: m6, and blue: m7) from \cite{V_2026}. We show the two best-fit dust evolution models from \cite{Calura_2023} (Model 5': blue solid; Model 9': maroon solid).}
    \label{fig:dgr_z_mixed}
\end{figure}

In this section, we discuss the resolved DGR--$Z$ and DMR--$Z$ relations in the context of previous observational surveys and dust evolution models. 
Our analysis emphasizes that the inferred correlations depend on the assumed molecular gas mass.

In particular, we find that the constant Galactic $\alpha_{\rm CO}$ (B13) is insufficient to describe the diverse ISM conditions observed in our spatially resolved analysis ($\sim$0.6 -- 2.3 kpc).
The diversity in $L_{\rm CO(1-0)}$--SFR ratios observed in our sample (see Fig.~\ref{fig:lco_sfr}), combined with information on metallicity and other ISM properties, further confirms that a single $\alpha_{\rm CO}$ cannot universally represent the molecular gas reservoir.
This motivates the adoption of a hybrid approach for $\alpha_{\rm CO}$.

Based on the results and considerations presented in Sect.~\ref{sec:co-sfr},
we divide our galaxies into two regimes:
\textit{i) CO-bright regime}: for galaxies exhibiting a typical ratio of $\rm \sim 8.1 \times 10^3 \, L_\odot (M_\odot \, \rm yr^{-1})^{-1}$ and with near-solar metallicity, we adopt the A16 prescription; \textit{ii) CO-dark regime}: for systems with ratios significantly below the above threshold and with sub-solar metallicity (specifically NGC~2403 and NGC~925), we adopt the S12 prescription. 
Although NGC~4736 falls within the CO-bright regime, the limited spatial coverage of its metallicity map (see Fig.~\ref{fig:metallicity_coverage}) prevents a robust derivation of a metallicity-dependent $\text{H}_2$ distribution; consequently, we opt for the B13 prescription for this target. While the transition between the CO-bright and CO-dark regimes is expected to be continuous, our classification captures the primary systematic differences driven by the local radiation field and dust shielding across the sample.

Before presenting the results for the DGR--Z and DMR--Z relations using the hybrid \(\alpha_{\rm CO}\) approach, it is important to clarify the key assumption underlying this methodology. Following the empirical $L_{\rm CO(1-0)}$-SFR relation of \cite{Bonato_2024}, we interpret systematic deviations from this relation mainly as arising from variations in $\alpha_{\rm CO}$, rather than from variations in $\tau_{\rm H_2}$. In this framework, galaxies that are underluminous in CO for their SFR are interpreted as hosting a larger fraction of CO-dark molecular gas, requiring a higher \(\alpha_{\rm CO}\). We acknowledge that variations in $\tau_{\rm H_2}$ can also contribute to the observed scatter in the $L_{\rm CO}$--SFR relation; however, our adopted methodology assumes that the dominant driver of the systematic offsets observed in our sample is $\alpha_{\rm CO}$ itself. The hybrid prescription should therefore be interpreted within this assumption.

When comparing our results with previous observational studies, it is also important to ensure consistency in the adopted metallicity calibrations in addition to the prescriptions for $\alpha_{\rm CO}$.
In this work, we adopt the \citet[][PG16S]{Pilyugin_2016} strong-line metallicity calibration, which is likewise used for most of the spatially resolved and global datasets selected for comparison in the following section (Figs.~\ref{fig:dgr_z_mixed} and \ref{fig:dmr_z_mixed}). An exception is \citet{RemyRuyer_2014}, who utilize the \citet[][PT05]{PT05} calibration, which may introduce a systematic offset in the derived metallicities. Nevertheless, this discrepancy does not alter the qualitative trends discussed here.
A similar consideration applies to $\alpha_{\rm CO}$.
Variations in the assumed functional form of $\alpha_{\rm CO}$ can systematically influence both the normalization and the slope of the DGR--$Z$ and DMR--$Z$ relations (as seen in Figs.~\ref{fig:dgr-Z} and \ref{fig:dmr-Z}). However, the majority of studies used here for comparison adopt a fixed MW value, for example, from B13 or similar works. Our hybrid methodology for determining $\alpha_{\rm CO}$ may therefore introduce an offset relative to studies that assume a constant conversion factor. In the following sections, we analyze the resolved DGR--$Z$ and DMR--$Z$ relations with these caveats in mind.

\subsection{Interpreting the DGR–Z relation: comparisons and underlying assumptions}
\label{sec: discussion-dgr}

In Fig.~\ref{fig:dgr_z_mixed},  we show the DGR--$Z$ relation for our sample using the hybrid \(\alpha_{\rm CO}\) prescription. 
The adoption of this approach yields a tighter DGR--$Z$ relation compared to the three individual prescriptions (Fig.~\ref{fig:dgr-Z}). 
In general, our resolved measurements are now better aligned with the \citet{DeVis_2019} curves and are closer to the trend reported by \citet{Galliano_2021}. 
In particular, the hybrid prescription reduces the scatter at the low-metallicity end by bringing galaxies such as NGC~2403 and NGC~925 closer to the main distribution.

Our resolved DGR--$Z$ relation is broadly consistent with studies of the MW disk. 
\citet{Giannetti_2017} investigated the variation of the DGR across the Galaxy, finding a power law dependence of DGR on metallicity as ${\rm DGR} \propto Z^{1.4}$. More recent determinations suggest a somewhat flatter dependence, ${\rm DGR} \propto Z^{1-1.2}$ (Giannetti et al., in prep.), indicating a reduced sensitivity of the DGR to metallicity at the metal-rich end, and show better agreement with our observations.

Our results are also consistent with DGR--$Z$ trends established through both global and resolved measurements in nearby galaxies.
Regarding global studies, in addition to \citet{DeVis_2019} and \citet{Galliano_2021}, our findings align well with the high-metallicity branch ($12 + \log(\rm O/H) > 8.1$) of the power law from \citet{RemyRuyer_2014}, shown as the green dashed line in Fig.~\ref{fig:dgr_z_mixed}. 
In this regime, the DGR is expected to scale almost linearly with metallicity. 
This is typical of galaxies where dust evolution is dominated by efficient grain growth in the ISM rather than primary stellar injection, as predicted by chemical evolution models \citep[e.g.,][]{Asano_2013, Zhuko_2016}.

Regarding resolved studies, \cite{Vilchez_2019} found a broken DGR--$Z$ relation in NGC~5457 and NGC~628 at the scale of $\sim$1.3~kpc, with a steep slope where $12+\log(\text{O/H}) < 8.4$ followed by a much shallower trend at higher metallicity.
In our sample, the limited spatial coverage of the metallicity map prevents us from probing the outer regions of NGC~5457 where this break becomes prominent; however, for the inner regions of NGC~5457, we recover a similar shallow decrease in DGR with metallicity, consistent with their observations at $\rm 12+log(O/H) > 8.4$. 
A complementary perspective is provided by \cite{Park_2024}, based on 11 TYPHOON galaxies, whose observed trends favor a broken power law description at the scale of \(0.2-2.3\) 
kpc (in Fig.~\ref{fig:dgr_z_mixed}, their derived relation is shown by the pink dot-dashed line).
While the overall trend with metallicity is broadly consistent, our measurements are systematically offset toward lower log(DGR) values by approximately 0.25 dex.
This offset could be driven by the choice of the dust model; our dust masses are derived using the full THEMIS model, which has been shown to provide dust masses lower by factors of 2--3 than other commonly used models \cite[e.g.,][]{Draine_2007}, as discussed in \citet{Casasola_2020} and \citet{Pas_2026}.
Beyond the absolute normalization, the slope of the relation can also be affected by dust modeling.
Our results are consistent with the recent resolved analysis of the face-on spiral M~99 by \citet{Pantoni_2026}, who found a nearly flat and scattered DGR--$Z$ trend at a scale of $\sim$1.8~kpc, similar to the behavior we observe across our sample and, in particular, within individual galaxies.  
\citet{Pantoni_2026} interpreted this behavior as closely linked to the choice of the dust model: while standard Modified Black Body fits—where the dust emissivity index (the parameter describing the wavelength dependence of dust emissivity) is often treated as a free parameter—tend to produce an increasing trend of DGR with metallicity, the THEMIS model favors a much flatter relation. 
This suggests that the shallow DGR--$Z$ trend we find could be, at least in part, a systematic feature of the adopted dust model.

We also compare our measurements with the cosmological dust-evolution simulation of \cite{tray_2025}, which models galaxy formation in a \(25^3\,{\rm cMpc}^3\) cosmological volume while self-consistently tracking dust production, ISM grain growth, and dust destruction. 
Their fiducial model reproduces an increasing DGR--$Z$ trend for star-forming galaxies. 
Fig.~\ref{fig:dgr_z_mixed} shows the median DGR--$Z$ relation from their Fiducial run (for galaxies with \(M_* > 10^9 \rm M_{\odot}\), brown dotted line). While the slope is broadly consistent, our observations are offset by \(\sim 0.5\) dex toward higher DGR values at fixed metallicity. 
This offset is driven by differences in metallicity definitions between simulations (e.g., gas-phase oxygen fractions in cold, dense gas) and observations (nebular abundance), as well as degeneracies in absolute metallicity scales (see their Appendix~C). This suggests that the normalization difference is driven by calibration rather than fundamental discrepancies in the underlying dust evolution.
In addition, our observations are compared with the fiducial model of \cite{Parente_2026}, which tracks the dust mass and grain size distribution evolution
in a semi-analytic cosmological framework that self-consistently accounts for stellar dust production, grain growth, and destruction processes.
In Fig.~5, we show the DGR–$Z$ relation from their fiducial run (red solid line).
The model reproduces an increasing DGR–$Z$ trend and is in good agreement with the bulk of our resolved measurements, although some galaxies show noticeable deviations.
Within their fiducial model framework, ISM grain growth via accretion emerges as the dominant process driving variations in the DGR–$Z$ relation (see their Figs.~7 and 8, left panel); suppressing this process results in a significantly flatter DGR–$Z$ relation and an overall deficit in dust mass.

Figure Fig.~\ref{fig:dgr_z_mixed} also shows the predictions from the chemical models of \cite{Calura_2023}. A dedicated discussion is deferred to Section~\ref{sec:models}.

In addition to the metallicity dependence, we investigate the relation between DGR and other local galaxy properties, such as $\Sigma_{\rm M_*}$ and the specific star formation rate (\(\rm sSFR = \frac{\Sigma_{\rm SFR}}{\Sigma_{M_*}}\)). Since the primary goal of this paper is to investigate how the DGR and DMR scales with metallicity, we defer the analysis of additional scaling relations to Appendix~\ref{appendix:relations}.  
We find a positive correlation between DGR and $\Sigma_{M_*}$,while DGR decreases with increasing sSFR. However, these trends show only moderate correlation strengths, with Pearson coefficients of \( r = 0.4\) and \(-0.5\), respectively.

\subsection{Interpreting the DMR--Z relation: comparisons and physical drivers}
\label{sec:interpret-dmr-z}

\begin{figure}[h]
    \centering
    \includegraphics[width=1\linewidth]{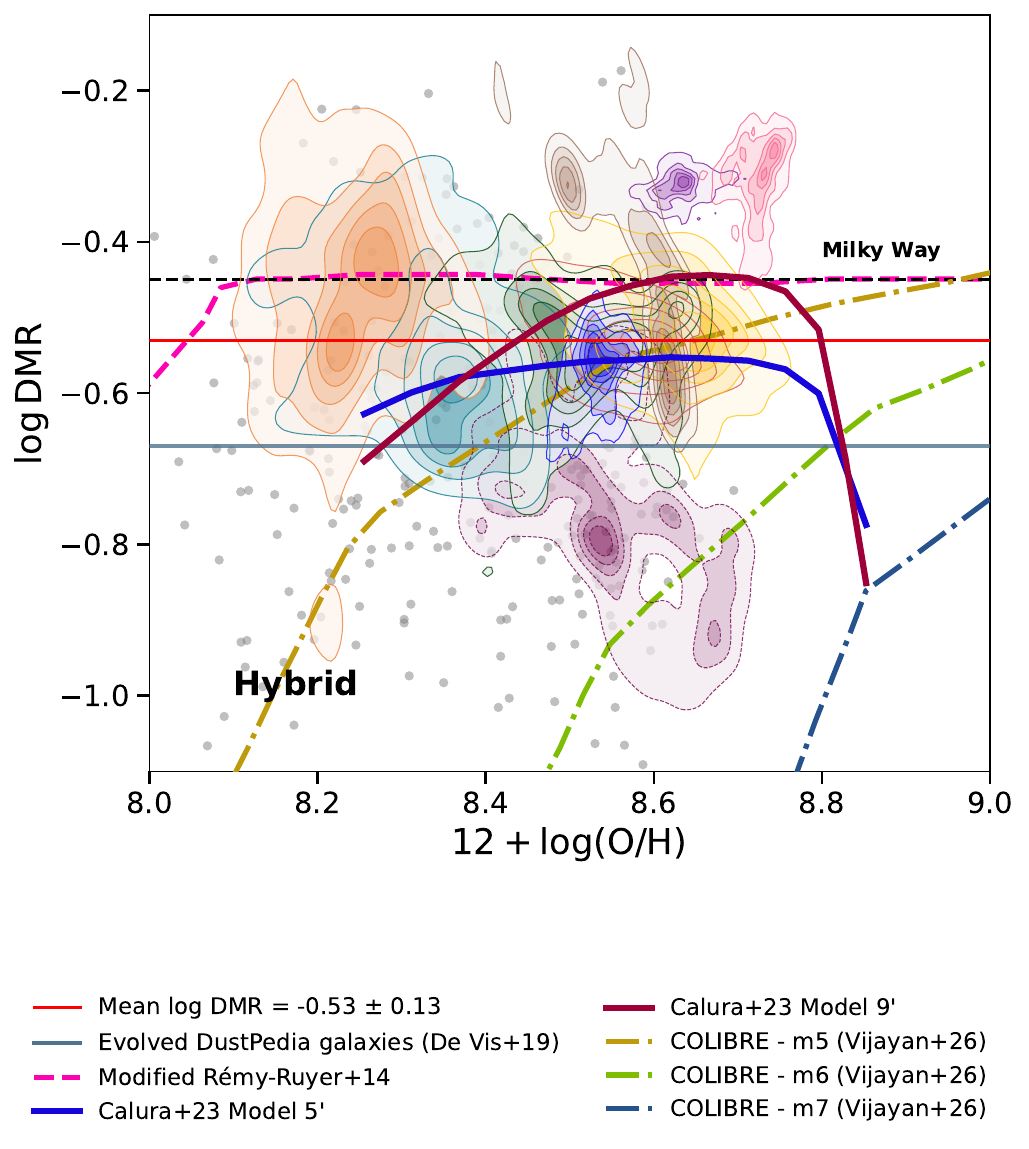}
    \caption{Resolved DMR as a function of gas-phase metallicity for our galaxy sample, computed using the hybrid \(\alpha_{\rm CO}\) prescription. Color of density distribution of each galaxy is same as in Fig.~\ref{fig:dgr-Z}. The black dashed line marks the MW DMR (\(\rm log DMR \sim -0.45\)), while the red solid line indicates the mean DMR of our sample (\(\rm log DMR \sim -0.53\)). The pink dashed line shows the high-metallicity DMR from \cite{RemyRuyer_2014}, and the gray solid line denotes the average DMR inferred for evolved DustPedia galaxies with \(\rm 12 + log(O/H) \geq 8.2 \) from \cite{DeVis_2019}. We also show the different prediction of COLIBRE simulation in dash-dot lines (yellow: m5 , greem: m6, and blue: m7) from \cite{V_2026}. We include the two best-fit dust evolution model predictions from \cite{Calura_2023} (Model 5’: blue solid; Model 9’: maroon solid).}
    \label{fig:dmr_z_mixed}
\end{figure}

Figure~\ref{fig:dmr_z_mixed} shows the resolved DMR--$Z$ relation for our sample using the hybrid \(\alpha_{\rm CO}\). 
Using our hybrid approach, the DMR–$Z$ relation becomes significantly tighter and nearly flat, with the majority of galaxies now confined to a narrow range of \(-0.6 \lesssim \log({\rm DMR}) \lesssim -0.4\). The resulting distribution has a mean value of \(\log({\rm DMR}) = -0.53 \pm 0.13\), as indicated by the red solid line in Fig.~\ref{fig:dmr_z_mixed}.
Some objects deviate from the main distribution: IC~342 has the lowest DMR, especially in its inner regions, with mean $\log({\rm DMR}) \sim -0.8$, while NGC~5055 and NGC~3521 
exhibit higher DMR values, with mean $\log({\rm DMR}) \sim -0.33$, approaching the theoretical upper limit expected for the DMR in galaxies \citep[see Appendix A of][]{Palla_2024} and consistent with the maximum values inferred from dust depletion studies across a range of ISM environments \citep{Konstantopoulou_2022}.
Compared to the distributions obtained with the individual $\alpha_{\rm CO}$ prescriptions (Fig.~\ref{fig:dmr-Z}), the hybrid approach suppresses the more pronounced variations and recovers a relatively flat DMR, consistent with the high-metallicity plateau reported by \citet{DeVis_2019} (gray line in Fig.~\ref{fig:dmr_z_mixed}). Indeed, although our mean DMR is higher by $\sim 0.1$ dex, our measurements remain well within the intrinsic scatter of their relation ($\sim$1 dex for $12+\log({\rm O/H}) \gtrsim 8.2$). 

For comparison, the MW value is indicated by the black dashed horizontal line in Fig.~\ref{fig:dmr_z_mixed} at $\log({\rm DMR}) \sim -0.45$, derived assuming a total gas mass of $M_{\rm gas} = 12.5 \times 10^9 \rm M_{\odot}$ \citep{Kalberla_2009}, solar metallicity, and a dust-to-HI ratio of 1/135 from the THEMIS model. Our results are broadly consistent with this value, although the mean DMR of our sample lies slightly below the Galactic estimate.

The absence of a pronounced DMR–$Z$ trend in our sample is consistent with previous large-scale surveys of nearby galaxies.
\citet{DeLooze_2020} showed that, while an increase of DMR with metallicity is observed when including low-metallicity, gas-rich systems, the relation flattens significantly for more massive ($M_* > 10^9 \rm M_{\odot}$) and evolved galaxies with $12 + \log(\mathrm{O/H}) \gtrsim 8.2$. In this regime, our measurements are consistent with an approximately constant DMR and fall within the intrinsic scatter of their results.
\citet{DeLooze_2020} further found that the DMR correlates more strongly with global galaxy properties such as stellar mass and gas fraction, suggesting that it primarily traces the overall evolutionary state of a galaxy rather than metallicity alone.

This evolutionary framework is also consistent with the flattening of the mass-metallicity relation observed in the high-mass regime (\(\rm \log(M/M_{\odot} > 10.5)\)) as a function of the stellar-to-gas mass ratio \citep{Zahid_2014}. In our sample, NGC~5055 and NGC~3521 are the most massive galaxies (\(\rm \log(M/M_{\odot} > 10.7)\)) and exhibit the highest DMR values, approaching the theoretical saturation limit. In this massive regime, past mergers and AGN feedback could have played a major role in shaping the ISM. Both systems are classified as LINERs, providing observational support for nuclear activity or shock excitation that could drive the observed local deviations in their ISM properties. 

The observed stability of the DMR against metallicity is further supported by the resolved analysis of \citet{Casasola_2022}, who found no clear trend across a similar metallicity range, suggesting that the critical threshold for efficient grain growth has already been surpassed.
However, other resolved studies also demonstrate that variations in the DMR at fixed metallicity can arise from local ISM conditions rather than metal enrichment alone. 
For example, in NGC~5457, \citet{Chiang_2018} demonstrated that, after removing radial trends, the DMR correlates with the molecular gas fraction, indicating a direct link between the dense molecular phase and the fraction of metals locked into dust.
\citet{Chiang_2021} further showed that inferred DMR values are sensitive to the adopted $\alpha_{\rm CO}$, a factor we explicitly address through our hybrid prescription. 
Independent evidence of this environmental dependence comes from the Magellanic Clouds, where \citet{RD_2017} found that the DGR increases by a factor of 3 in the Large Magellanic Clouds and 7 in the Small Magellanic Clouds when transitioning from the diffuse to the dense ISM. 
Subsequent depletion-based measurements further indicate that the fraction of metals locked into dust rises with increasing hydrogen column density \citep{Roman_Duval_2021, Roman_2022}, identifying local gas density as the main factor driving variations in the DMR within galactic disks. 

The role of environment may also contribute to the scatter observed in our sample. Given the regular spiral morphologies of our galaxies, they are unlikely to have experienced recent major mergers. However, as discussed in \citet{Tailor_2025}, almost all these systems are classified as members of galaxy groups with varying degrees of tidal interactions. While these minor interactions are not strong enough to disrupt the stellar disk, they can introduce localized perturbations in the ISM conditions \citep[e.g.,][]{Casasola_2004}, potentially contributing to the observed internal scatter in the ISM scaling relations.

The behavior of the DMR--$Z$ relation is also influenced by the assumed functional form of the $\alpha_{\rm CO}$. When applying single, steep metallicity-dependent prescriptions (such as S12) across an entire sample, negative DMR--$Z$ trends can emerge. This is partly an artifact of mathematical coupling, as metallicity appears in the denominator of the DMR definition (see Sect.~\ref{sec:ratios}).  Applying a steep $\alpha_{\text{CO}}(Z)$ prescription increases the inferred molecular gas mass in lower-metallicity regimes, thereby lowering the DGR. In turn, this depresses the DMR at low metallicity, artificially introducing a stronger metallicity dependence in the DMR--$Z$ relation. Our hybrid approach mitigates this effect by adopting the S12 calibration only where physically motivated by CO photodissociation and retaining more standard prescriptions elsewhere. This prevents artificial slopes and yields a flat DMR--$Z$ relation consistent with physical expectations.

The nearly-constant DMR observed in our sample can be understood as a natural consequence of dust evolution in the metal-rich regime.
The DMR represents the balance between dust formation/growth and destruction mechanisms in the ISM. 
Dust evolution models predict a transition from low DMR values at early stages (when dust production is dominated by stellar sources) to a rapid increase once grain growth in the ISM becomes efficient, followed by a saturation at high metallicities \citep[e.g.,][]{Asano_2013,Zhuko_2016, Gioannini_2017,Galliano_2018, Palla_2024}. This flattening arises because dust formation becomes limited by stoichiometric constraints which imposes an upper bound on how many metals can be incorporated into dust grains, as most available species are progressively locked into solids \citep[see][Appendix A]{Palla_2024}.
Although our data do not probe the low-metallicity regime where this transition occurs, the absence of a strong DMR–$Z$ trend and the nearly constant value of DMR suggest that our sample lies within this evolved, grain growth-dominated phase of dust evolution. 
In this sense, our results provide resolved-scale support for the high-metallicity plateau inferred from global studies.

When comparing our resolved observations to dust-evolution models, it is important to distinguish between spatially integrated single-zone models and high-resolution cosmological simulations.
Classic single-zone semi-analytical frameworks treat galaxies as single systems, where different metallicities primarily reflect different global evolutionary stages. In contrast, high-resolution cosmological simulations can resolve individual galaxies into multiple spatially distinct sub-components,  allowing for variation in metallicity and dust properties within each galaxy.
High-resolution cosmological simulation produces substantial spatial variations in the DMR within individual galaxies, including central regions with reduced DMR associated with intense star formation and efficient supernova-driven dust destruction \citep[e.g.,][]{Byun_2025, tray_2025}. This provides a complementary theoretical perspective to our  resolved observations, in which higher metallicity primarily traces the inner, denser regions of optical disks due to radial metallicity gradients.
On these local scales, processes such as local dust destruction by supernovae, variations in star formation efficiency, and ISM inhomogeneities can therefore introduce scatter around the equilibrium DMR plateau. To further investigate such small-scale variations, we examined the DMR–Z relation using radial-profile averages instead of individual spatial pixels (see Appendix~\ref{appendix:radial_DMR}). In this case, the relation becomes flatter, demonstrating that part of the internal scatter observed in our pixel-by-pixel analysis arises from local ISM variations that are averaged out when using radial profiles.

For completeness, we also examine the DMR relations with $\Sigma_{M_*}$ and sSFR (Appendix~\ref{appendix:relations}). Unlike the DGR, the DMR shows no dependence on $\Sigma_{M_*}$ and only a weak anti-correlation with sSFR. This supports the scenario where the fraction of metals locked into dust grains remains stable across different local environments.

\subsection{Comparison with dust evolution models for spiral galaxies}
\label{sec:models}

A detailed galaxy-by-galaxy modeling of the dust and metal evolution is beyond the scope of this work. 
Nonetheless, we adopted the \cite{Calura_2023} (hereafter C23) models, which are calibrated to the nearby spiral galaxy NGC~628, part of our sample.
Given that NGC~628 is a typical, metal-rich, star-forming spiral, its evolutionary tracks provide a useful reference for the bulk of galaxies in our sample. 
However, it is important to note that these models are specifically tuned to reproduce the specific properties of NGC~628 and, therefore, their applicability is limited to systems with similar ISM properties. 
Galaxies that differ significantly in terms of gas content, star formation activity, and stellar masses may not be fully described by this framework, and deviations from the model predictions should be interpreted with caution.

C23 developed detailed multi-zone chemical evolution models for NGC~628, using an MCMC algorithm to fit star formation and chemical evolution parameters to the stellar and gas radial profiles presented in \citet{Morselli_2020}, allowing an in-depth parametric study of the effect of dust parameters on dust radial profile.
In C23 dust evolution analysis, they varied two main quantities: (i) the recipe for dust accretion time scale 
and (ii) mass of gas cleared out of dust by a supernova (\(\rm{M_{clear}}\)).
They found two model solutions that fit the observed dust radial profile well (Model 5$^\prime$ and Model 9$^\prime$). For full details on the model prescriptions, we refer the reader to C23. 

In both models, stellar dust production (AGB + SN; characterized by a timescale \(\tau_{*}\)) governs the dust budget with respect to grain growth (with timescale \(\tau_g\)) at early times, with a comparable \(\tau_{*}/\tau_g\) ratio up to 5 Gyr. After 5 Gyr, grain growth becomes the dominant dust source, with Model 5$^\prime$ exhibiting a higher dust growth rate relative to Model 9$^\prime$.
The two models differ significantly in how the dust destruction timescale (\(\tau_{d}\)) compares to \(\tau_g\) and in how this ratio evolves over time.
Model 5$^\prime$ attains equilibrium in the \(\tau_g/\tau_d\) ratio after 8 Gyr and then stays constant up to the present, with \(\tau_g / \tau_d \approx 1\).
For Model 9$^\prime$, this ratio remains approximately constant at \(\tau_g / \tau_d \approx 0.3\) from around 8 Gyr to the present time, thus, for this model, the ISM grain growth dominates over destruction (see Fig. 14 in C23).
Here, we compare our measurements with the present-day evolutionary tracks of the two best-fit models (5$^\prime$: blue, 9$^\prime$: maroon), focusing on how well each model reproduces the bulk distribution of our galaxy sample. 

In terms of the DGR--$Z$ relation (Fig.~\ref{fig:dgr_z_mixed}), both Model 5$^\prime$ and Model 9$^\prime$ reproduce the observed increase of DGR with metallicity across the majority of galaxies, indicating that both parameterisations provide a reasonable description of the dust content in these star-forming systems. The two models differ in slope, with Model 9$^\prime$ predicting a steeper increase of DGR with metallicity and providing a slightly better visual agreement with the distribution at the highest metallicities.
To quantify the agreement, we compute the residuals between the observed DGR values and each model prediction, and the resulting root mean square error (RMSE) is very similar for the two models (0.29 dex for Model 5$^\prime$ and 0.30 dex for Model 9$^\prime$)
indicating no statistically significant improvement of one model over the other. Therefore, while both models successfully reproduce the overall DGR--$Z$ trend, the DGR relation alone is not sufficient to break the degeneracy between dust growth and destruction.

The DMR--$Z$ relation (Fig.~\ref{fig:dmr_z_mixed}) shows that the observational data are broadly consistent with a roughly constant DMR over the metallicity range probed, with values clustering around $\log(\mathrm{DMR}) \approx -0.53$ and no strong systematic dependence on $Z$. Model 5$^\prime$ reproduces this behavior well, yielding a nearly flat DMR--$Z$ relation consistent with the observed level. Model 9$^\prime$ instead predicts a mild curvature, with a weak variation at intermediate metallicities followed by a decline toward higher $Z$. Similar to DGR--$Z$ case, to quantify the agreement, we compute the residuals between the observed DMR values and each model prediction, and resulting RMSE is nevertheless very similar for the two models (0.14 dex for Model 5$^\prime$ and 0.16 dex for Model 9$^\prime$), indicating comparable statistical performance despite their different functional forms.

These results should therefore be interpreted in light of the calibration limitations discussed before. 
While both models provide a comparable global level of agreement with the data in terms of RMSE, this does not necessarily imply uniform validity across the full sample, given that the frameworks are anchored to a single reference system. 
The comparable statistical performance of Model 5$^\prime$ and Model 9$^\prime$ demonstrates that present-day, sample-wide DGR--$Z$ and DMR--$Z$ relations alone are subject to a strong degeneracy, and different evolutionary pathways can successfully sustain the same observed scaling relations. 
Breaking this degeneracy and understanding the observed galaxy-to-galaxy deviations will require dedicated, galaxy-specific modeling that accounts for local variations in star formation histories, gas fractions, and dust grain size distributions, rather than relying on a single global prescription.

\subsection{Deviations from the mean DGR and DMR: environmental and systematic effects}
\label{sec:deviations}

Although most of the systems in our sample exhibit relatively uniform DGR and DMR, a small subset presents systematic deviations in both quantities. 
In particular, NGC~5055 and NGC~3521 display elevated DGR and DMR values compared to the bulk of the sample, whereas IC~342 lies at the lower end of the range for both ratios (see Fig.~\ref{fig:dgr_z_mixed} and ~\ref{fig:dmr_z_mixed}). 
These deviations indicate differences in the efficiency with which metals are locked into dust, suggesting variations in the local ISM environment and/or systematic uncertainties in the inferred gas and dust masses.

From a global perspective, the elevated DMR and DGR values in NGC~5055 and NGC~3521 are consistent with differences in evolutionary stages.
Studies of nearby galaxies indicate that the DMR is more tightly correlated with the gas fraction ($f_{\rm gas} = M_{\rm gas} / (M_{\rm gas} + M_*)$), which serves as a proxy for the evolutionary stage of a galaxy. \cite{DeVis_2019} found that the more evolved galaxies (i.e., those with lower $f_{\rm gas}$) tend to exhibit higher DMR values, consistent with a longer timescale available for the conversion of gas-phase metals into dust, combined with a reduced rate of dust destruction due to SNe shocks associated with their typically lower specific star formation rates.
This finding is further reinforced by \citet{Casasola_2022}, who observed a similar DMR--\(f_{\rm gas}\) trend using spatially resolved data, suggesting that this link between ISM evolution and dust enrichment holds from global to local scales.
In our sample, NGC~5055 and NGC~3521 indeed show systematically lower gas fractions, with median values of $f_{\rm gas} \sim 0.04$ and $\sim 0.02$, respectively, compared to typical values of $\sim 0.1$ – 0.4 for the rest of the galaxies.
This evidence strengthens the interpretation that the higher DMR values observed in these systems are driven by their advanced evolutionary state, indicating a more processed ISM relative to the rest of the sample, where grain growth has proceeded more effectively.

At the same time, as already said in Sect.~\ref{sec:co-sfr}, using either a constant or purely metallicity-dependent \(\alpha_{\rm CO}\) may fail to capture the full complexity of the ISM in all galaxies in our sample. This is proven for IC~342 \citep[][]{Chiang_2021}, but it could be extended to other galaxies.
Indeed, differences in gas surface density, temperature, dynamical state, the UV radiation field, cosmic ray density, and the overall local environment can lead to deviations from the adopted $\alpha_{\rm CO}$ \citep[e.g.,][]{Maloney_1988,Boselli_2002,Casasola_2007,Bolatto_2013}. The particularly low DMR observed in the central regions of IC~342 may also be related to local dust destruction. \citet{Byun_2025} found that intense central star formation can produce localized reductions in the DMR through efficient supernova-driven dust destruction, resulting in a central cavity. Although their study focuses on high-redshift galaxies and is not directly comparable to IC~342, this provides a possible physical explanation for the low central DMR, in addition to the systematic uncertainties.

As already mentioned, regarding the impact of dust SED modeling, additional uncertainties could arise from assumptions about dust properties. 
The inferred dust mass also depends on the adopted dust mass absorption coefficient, \(\kappa\), which is not expected to be uniform across different environments \citep[e.g.,][]{Bianchi_2019,Clark_2019,Pozzi_20}.
Moreover, variations in grain properties, such as coagulation and the accretion of mantles in dense regions, can increase the FIR emissivity of the grains \citep{K_2015,Ysard_2015,Jones_2017}. 
As a result, assuming a constant \(\kappa\) may lead to systematic overestimations in dust masses, and consequently to systematic offsets in DGR and DMR.

Taken together, these considerations suggest that the observed deviations in NGC~5055, NGC~3521 and IC~342 likely reflect a combination of environmental differences and uncertainties associated with the conversion from observables to gas and dust masses.
These galaxies serve as important test cases for dust and gas modeling, demonstrating that a one-size-fits-all approach to conversion factors may be insufficient for systems with extreme ISM properties or starburst activity. 
Consequently, these outliers not only represent scatter, but rather provide crucial insights into the diverse physical processes governing the life cycle of the ISM in nearby galaxies.

 \section{Conclusions}
 \label{sec:conclusions}
In this study, we present a spatially resolved analysis of the DGR–$Z$ and DMR–$Z$ relations across a sample of 10 nearby spiral galaxies, with stellar masses spanning \(9.7 \le \log (M_*/M_{\odot}) \le 11.0\), SFR of \(\sim\) 0.3–3 \(M_{\odot}\,\text{yr}^{-1}\), metallicities \(8.1 \leq 12 + \log(\rm O/H) \leq 8.8\), and covering physical scales ranging from $\sim$0.6 to 2.3 kpc.
We explore how the inferred relations depend on the adopted $\alpha_{\rm CO}$ by comparing three commonly used prescriptions: a constant Galactic value \citep{Bolatto_2013} and two metallicity-dependent functions, specifically \citet[$\alpha_{\rm CO} \propto Z^{-1.5}$]{Amorin_2016} and \citet[$\alpha_{\rm CO} \propto Z^{-2}$]{Schruba_2012}), together with a hybrid approach for $\alpha_{\rm CO}$. 

Our main results are as follows:
 \begin{itemize}
     \item Both DGR and DMR are sensitive to the adopted \(\alpha_{\rm CO}\), with different prescriptions showing systematic offsets and differences in the overall observed trends with respect to metallicity.
     \item Using the \cite{Bolatto_2013} prescription, the molecular gas masses of CO-faint and low-metallicity systems (NGC~2403 and NGC~925) are likely underestimated, leading to elevated DGR and DMR values. While this prescription provides values broadly consistent with previous studies for several metal-rich galaxies, it does not capture the diversity of molecular gas properties across our sample.
     \item The \cite{Amorin_2016} prescription introduces systematic shifts in the inferred DGR and DMR values. While the increased \(\alpha_{\rm CO}\) at low metallicity raises the molecular gas masses, the low-metallicity galaxies in our sample remain offset toward higher DGR and DMR values. At high metallicity, the lower \(\alpha_{\rm CO}\) values lead to higher DGR estimates in the inner regions of several galaxies. Overall, the \cite{Amorin_2016} prescription reduces some internal DMR gradients but does not significantly reduce the galaxy-to-galaxy scatter in the DGR--$Z$ and DMR--$Z$ relations.
     \item The \cite{Schruba_2012} prescription yields the largest molecular gas masses at low metallicity, shifting the DGR toward lower values, reducing the scatter in the DGR--$Z$ relation, and inducing a steeper metallicity dependence in the DMR--$Z$ relation.  
     \item We find that no single \(\alpha_{\rm CO}\) provides an adequate description of our entire sample. 
     Specifically, the constant Galactic \(\alpha_{\rm CO}\) proves to be too restrictive for our spatially resolved analysis.
     \item We adopt a hybrid $\alpha_{\mathrm{CO}}$ prescription combining the $L_{\mathrm{CO}}$--SFR diagnostic and metallicity to separate standard disks from CO-dark gas regimes. Using this approach, the DGR scales linearly with metallicity ($-2.75 \lesssim \log(\mathrm{DGR}) \lesssim -2.00$), aligning with observational and theoretical expectations. Conversely, the DMR remains roughly constant at $\log(\mathrm{DMR}) = -0.53 \pm 0.13$, indicating that $\sim 30\%$ of metals are locked into dust. This confirms a high-metallicity saturation plateau where ISM grain growth balances dust destruction.
     \item Local deviations from the mean DGR and DMR relations stem from both intrinsic disk physics and modeling uncertainties. Massive spirals (e.g., NGC~5055, NGC~3521) set the upper limit of the DMR saturation plateau, whereas the central starburst IC~342 is an outlier in the $L_{\rm CO}$--SFR plane owing to enhanced CO emissivity. These cases illustrate how starburst activity, evolutionary stage, and systematic assumptions drive scatter in resolved relations.
     \item We compare our observations with the two best-fit chemical and dust evolution models from \citet{Calura_2023}. Both models reproduce the overall DGR--$Z$ and DMR--$Z$ distributions of our sample with comparable statistical performance.
     \item Additional scaling relations reveal that while DGR correlates with $\Sigma_{M_*}$ and anti-correlates with sSFR, DMR is independent of $\Sigma_{M_*}$ and depends weakly on sSFR. This confirms that DMR is governed by the total metal budget rather than local disk properties.
     
 \end{itemize}

Our study highlights that high-resolution, spatially resolved maps of gas, dust, and metallicity are essential to capture the complex ISM physics that a single $\alpha_{\mathrm{CO}}$ cannot capture. Future progress requires extending these resolved analyses to lower-metallicity regimes to probe where grain growth becomes inefficient, better understand environmental and morphological dependencies, and place tighter constraints on dust evolution models.

\begin{acknowledgements}
The authors thank the anonymous referee for their constructive comments and valuable suggestions, which helped improve the quality and clarity of the manuscript.
VT, VC, FP, FC, and JF acknowledge funding from the INAF Mini Grant 2024 program `DustPedia meets Metal-THINGS: Dust-METAL'.
JF acknowledges financial support from the DGAPA-PAPIIT project IN102226, Mexico.
MALL  acknowledges grant RYC2020-029354-I funded by MICIU/AEI/10.13039/501100011033 by “ESF Investing in your future”, by “ESF+”
MALL, LEG, MMP, and 
GV acknowledges support from the Comunidad de Madrid through grant PIPF-2024/TEC-34555, and the Atracción de Talento program 2022-T1/TIC-23797, as well as grant PID2023-146372OB-I00 funded by MICIU/AEI/10.13039/501100011033 and by ERDF, EU.
SAC. (CVU 2054181) thanks SECIHTI for the MSc
scholarship and for the complementary mobility support provided through
the 2026 Convocatoria de Apoyos Complementarios de Movilidad en el
Extranjero, Movilidad Nacional, Movilidad en los Sectores de Interés y
Movilidad para Programas de Doble Titulación.   
VT is pleased to acknowledge the hospitality and stimulating environment provided by IRyA-UNAM, where part of the work on this paper was carried out during her stay in Morelia (Mexico). L.S.P acknowledges support from the Research Council of Lithuania (LMTLT) (grant no. P-LU-PAR-26-15). 
\end{acknowledgements}

\bibliographystyle{aa}
\bibliography{ref}
\clearpage

\begin{appendix}
\onecolumn
\section{Resolved distributions of dust, gas, DGR, and DMR for NGC~5457}
\label{appendix:ngc5457}

\begin{figure}[h]
    \centering
    \includegraphics[width=1\linewidth]{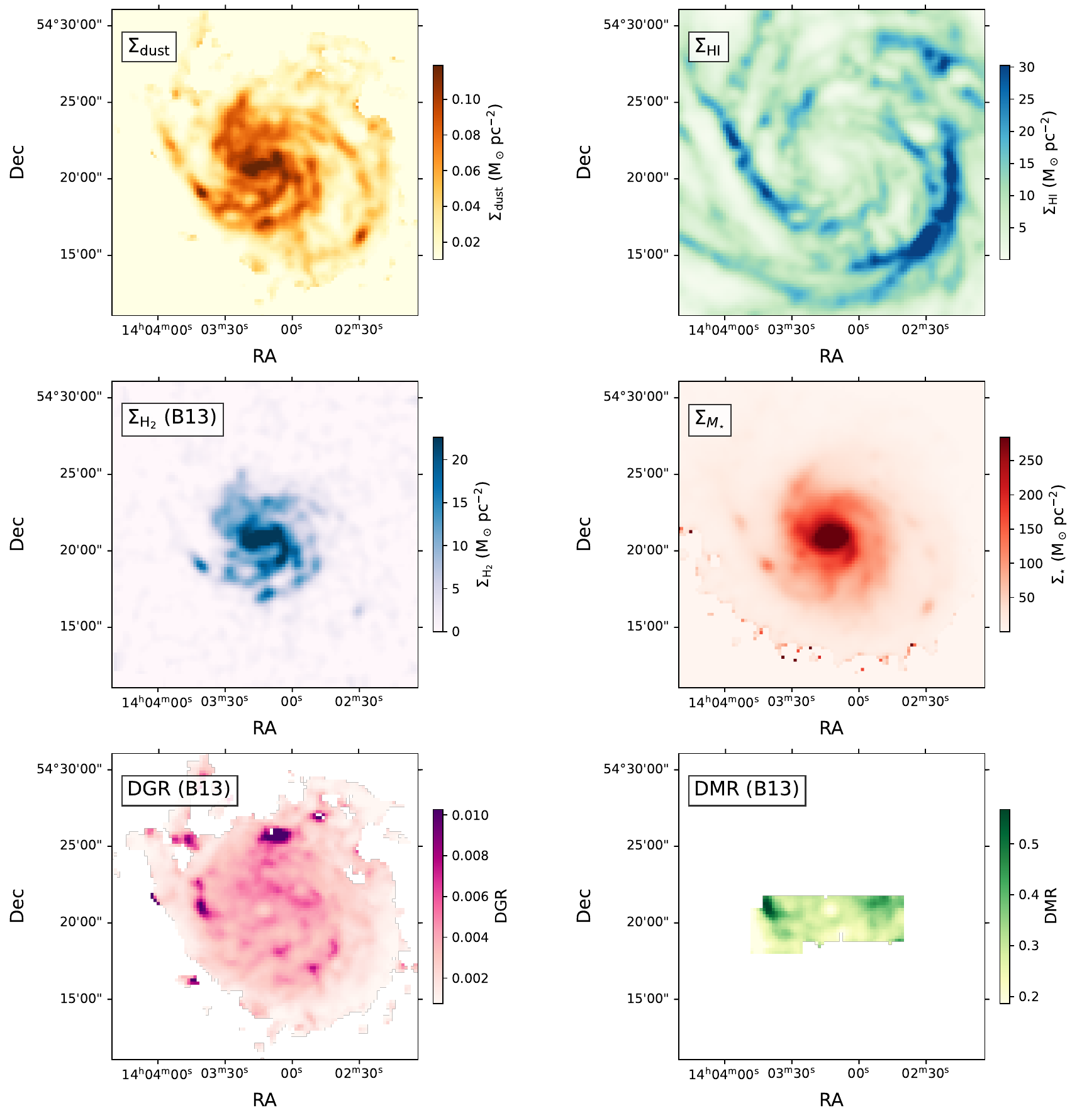}
    \caption{Example of the spatial distribution of the main ISM components alongside the derived DGR and DMR for NGC~5457. From left to right and top to bottom, the panels display $\Sigma_{\mathrm{dust}}$, $\Sigma_{\mathrm{H\,\textsc{i}}}$, $\Sigma_{\mathrm{H}_2}$ (derived using the B13 $\alpha_{\mathrm{CO}}$ prescription), $\Sigma_{M_*}$, DGR, and DMR.
    The maps are presented on a common spatial grid to illustrate the resolved distributions used throughout the DGR and DMR analysis.}
    
    \label{fig:example_NGC5457}
\end{figure}

In Fig.~\ref{fig:example_NGC5457}, we show, as an example, the matched-resolution and regridded maps of $\Sigma_{\mathrm{dust}}$, $\Sigma_{\mathrm{H\,\textsc{i}}}$, $\Sigma_{\mathrm{H}_2}$ (under the B13 prescription for $\alpha_{\mathrm{CO}}$), $\Sigma_{M_*}$, DGR, and DMR for the galaxy NGC~5457, following the procedures described in Sects.~\ref{sec:sample} and~\ref{sec:methods}.

\clearpage
\twocolumn
\section{Dependence of DGR and DMR on $\Sigma_{M_*}$ and sSFR}
\label{appendix:relations}

\begin{figure}[htbp]
    \centering
    \includegraphics[width=1\linewidth]{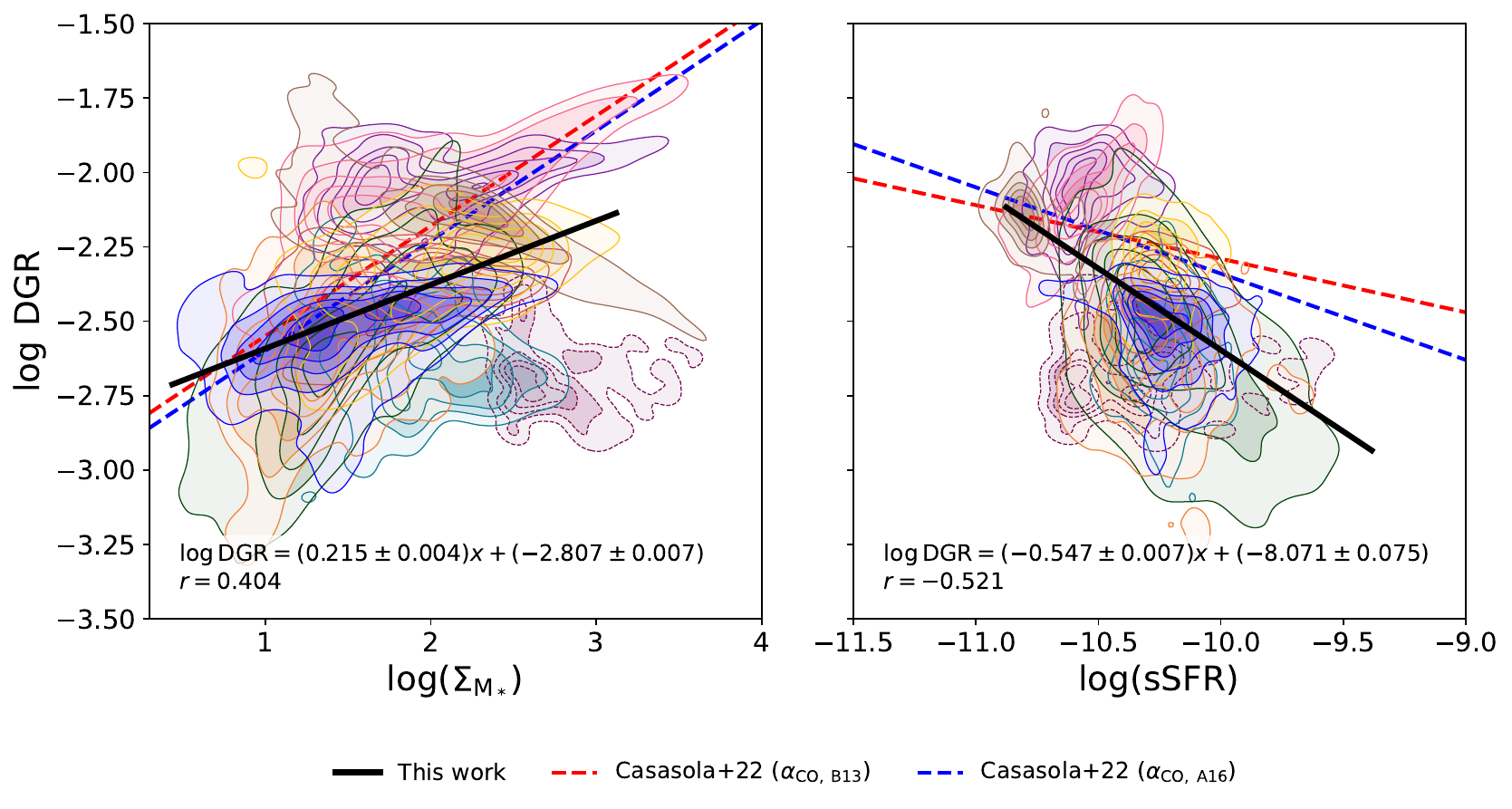}
    \caption{Resolved $\log(\mathrm{DGR})$ vs. $\log(\Sigma_{M_*})$ (left) and $\log(\mathrm{sSFR})$ (right) for our galaxy sample. Color contours match Fig.~\ref{fig:dgr-Z}. The black solid line shows our best-fit linear relation; each panel lists the equation and Pearson coefficient. Red and blue dashed lines show relations from \citet{Casasola_2022} using constant (B13) and metallicity-dependent (A16) $\alpha_{\mathrm{CO}}$, respectively.}
    
    \label{fig:DGR_properties}
\end{figure}

\begin{figure}[htbp]
    \centering
    \includegraphics[width=1\linewidth]{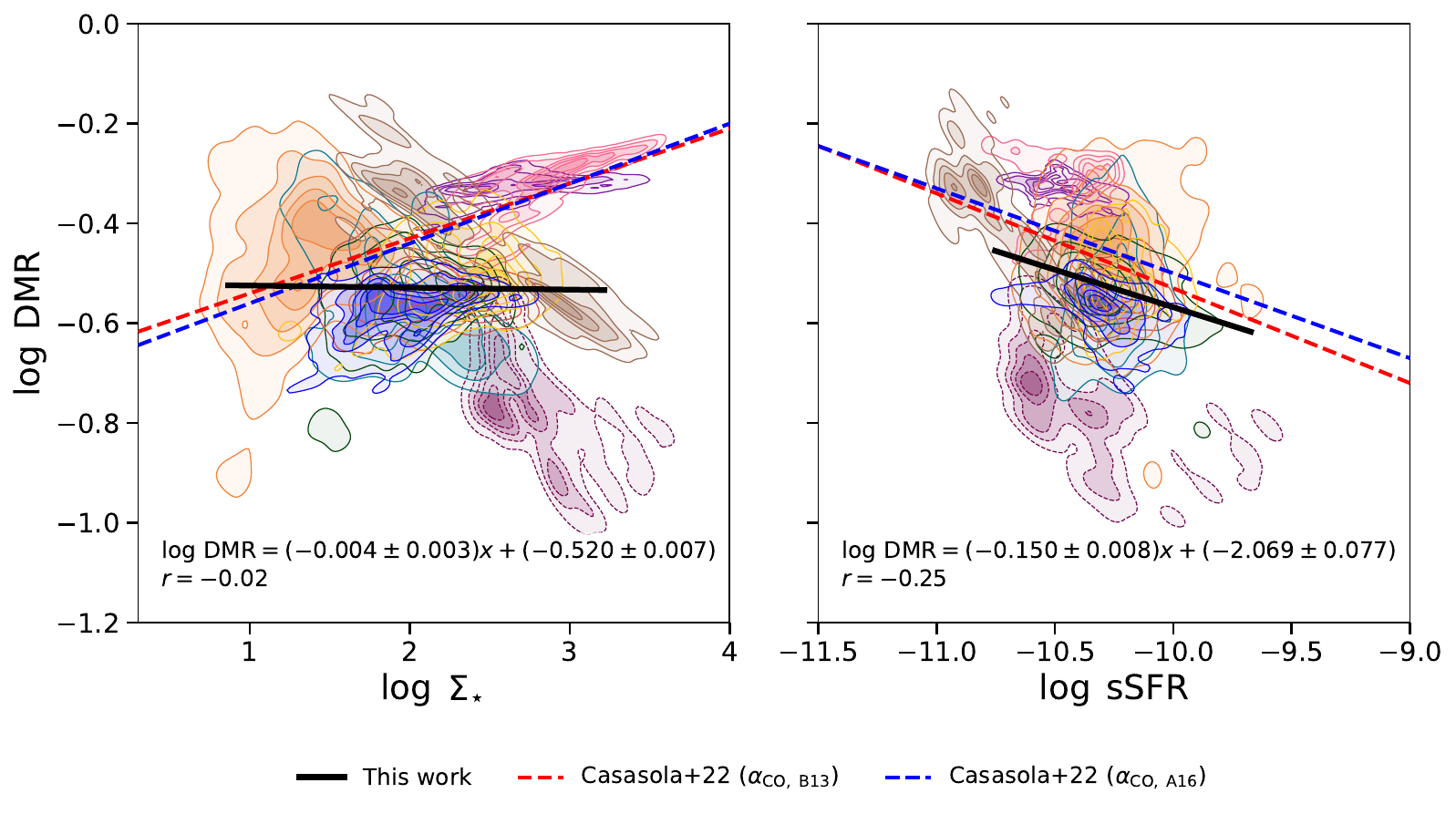}
    \caption{Same as Fig.~\ref{fig:DGR_properties}, but showing the resolved $\log(\mathrm{DMR})$ correlations with $\log(\Sigma_{M_*})$ (left) and $\log(\mathrm{sSFR})$ (right).}
    \label{fig:DMR_properties}
\end{figure}

We also investigated how the resolved DGR and DMR vary with local galaxy properties, specifically $\Sigma_{M_*}$ and sSFR. The results of this analysis are shown in Figs.~\ref{fig:DGR_properties} and \ref{fig:DMR_properties}, obtained using our hybrid $\alpha_{\mathrm{CO}}$ approach. For comparison, the corresponding resolved relations from \citet{Casasola_2022} are shown.

We find a positive correlation between DGR and $\Sigma_{M_*}$. For our sample, the best fit is $\log(\mathrm{DGR}) = (0.215 \pm 0.004)\log(\Sigma_{M_*}) + (-2.807 \pm 0.007)$ ($r=0.40$), consistent with trends in nearby galaxies where denser, more massive regions have higher DGR values \citep[e.g.,][]{DeLooze_2020}. 
However, our slope is shallower than \citet{Casasola_2022}, implying a weaker dependence on $\Sigma_{M_*}$. 

DGR is anti-correlated with sSFR, with $\log(\mathrm{DGR}) = (-0.547 \pm 0.007)\log(\mathrm{sSFR}) + (-8.071 \pm 0.075)$ ($r=0.52$). 
The opposite signs are expected because higher stellar mass systems generally have lower sSFR. 
This stronger anti-correlation than reported by \citet{Casasola_2022} likely reflects differences in how metallicity maps are derived (Sect.~\ref{sec:interpret-dmr-z}) and our hybrid $\alpha_{\mathrm{CO}}$ approach.

The DMR shows no significant dependence on $\Sigma_{M_*}$: $\log(\mathrm{DMR}) = (-0.004 \pm 0.003)\log(\Sigma_{M_*}) + (-0.520 \pm 0.007)$ with $r=-0.02$. The near-zero slope implies a roughly constant fraction of metals locked in dust across a wide range of $\Sigma_{M_*}$, so the DGR--$\Sigma_{M_*}$ variation is mainly driven by changes in gas-phase metallicity rather than by changes in dust production efficiency or metal depletion rates. DMR is only weakly anti-correlated with sSFR, $\log(\mathrm{DMR}) = (-0.150 \pm 0.008)\log(\mathrm{sSFR}) + (-2.069 \pm 0.077)$ ($r=-0.25$). Compared to the steeper DGR--sSFR relation, this shallow slope and weak correlation indicate that regions with differing star formation activity retain broadly similar fractions of metals in dust grains. This behavior is consistent with \citet{Casasola_2022}, who reported flatter DMR trends compared to DGR. The stability of the DMR against local $\Sigma_{M_*}$ and sSFR supports the conclusion that DGR variations across spiral disks are set by the total metal budget, whereas the efficiency of metal incorporation into dust remains broadly constant.

\section{Radially averaged DMR--$Z$ relation}
\label{appendix:radial_DMR}

\begin{figure}[h]
    \centering
    \includegraphics[width=0.8\linewidth]{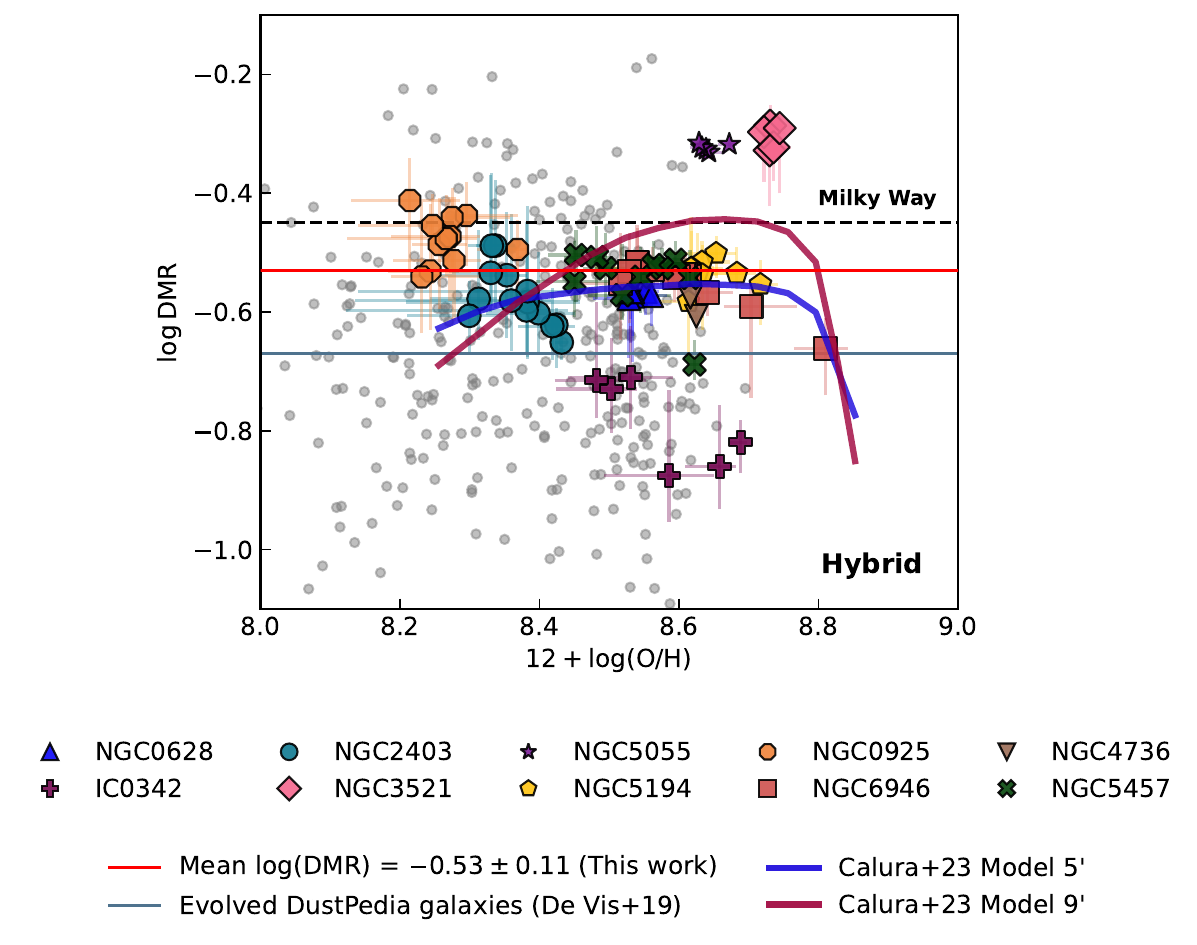}
    \caption{Radially averaged DMR--$Z$ relation for the galaxy sample using the hybrid $\alpha_{\mathrm{CO}}$ approach. Coloured symbols show median values within radial bins, with error bars indicating the 16th--84th percentile range. The black dashed line marks the MW DMR, the red solid line the sample mean, and the gray solid line the mean DMR of evolved DustPedia galaxies with $12+\log(O/H)\ge8.2$ \citep{DeVis_2019}. Blue and maroon solid curves show the best-fit dust-evolution models 5' and 9', respectively, from \citet{Calura_2023}.}
    \label{fig:radial_avg_dmr}
\end{figure}

To investigate the dependence of the DMR--$Z$ relation on spatial scale, we repeated the analysis using radial averages rather than individual pixels. Figure~\ref{fig:radial_avg_dmr} shows the resulting relation for the hybrid $\alpha_{\mathrm{CO}}$ approach, with the individual galaxies represented using the same symbols as in Fig.~\ref{fig:lco_sfr}. The colored points show the median DMR and $Z$ values in radial bins of width $24\arcsec$, corresponding to physical scales of $\sim0.4$--$1.5$~kpc, with error bars indicating the 16th--84th percentile range.

Radial averaging substantially reduces the scatter of the pixel-by-pixel relation, producing a tighter and nearly flat DMR--$Z$ relation. Most galaxies lie within $-0.6 \lesssim \log(\mathrm{DMR}) \lesssim -0.4$, although some remain offset. IC~342 reaches the lowest DMR values, with $\log(\mathrm{DMR}) \sim -0.8$, while NGC~5055 and NGC~3521 show higher values of $\sim-0.35$ and $\sim-0.30$, respectively. This indicates that part of the scatter in the pixel-by-pixel DMR--$Z$ relation is associated with local ISM variations that are smoothed out by radial averaging, highlighting the dependence of the observed scaling relation on spatial scale.

\onecolumn
\section{Radial Profiles for DGR}
\label{appendix: radial prof dgr}

\begin{figure}[htbp]
\centering
\begin{tabular}{cc}
\includegraphics[width=0.49\textwidth]{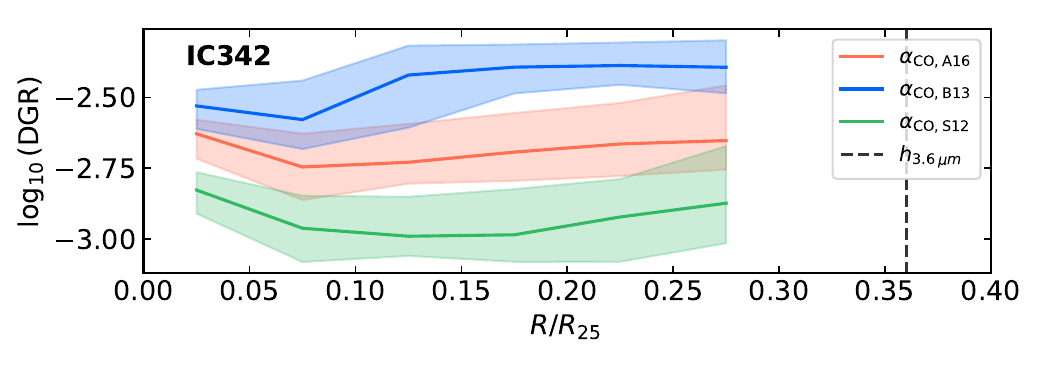} &
\includegraphics[width=0.49\textwidth]{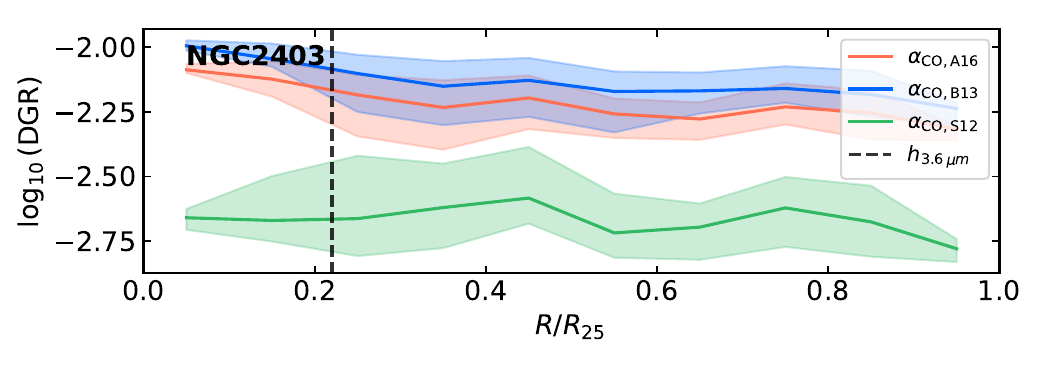} \\[1ex]
\includegraphics[width=0.49\textwidth]{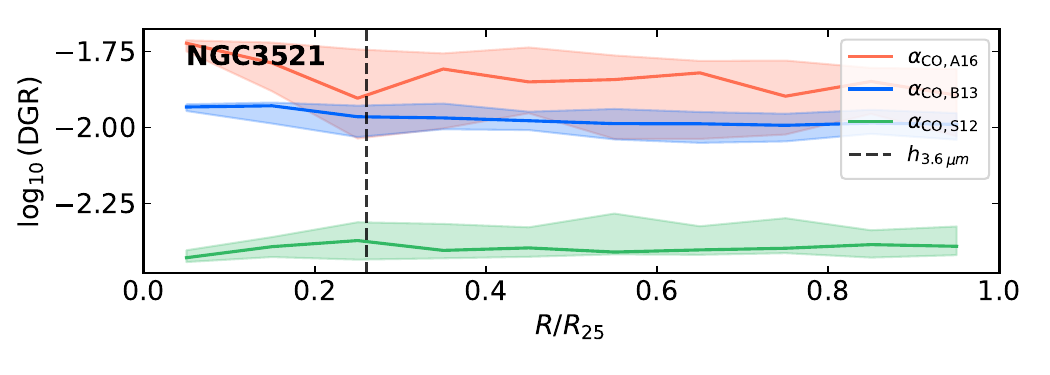} &
\includegraphics[width=0.49\textwidth]{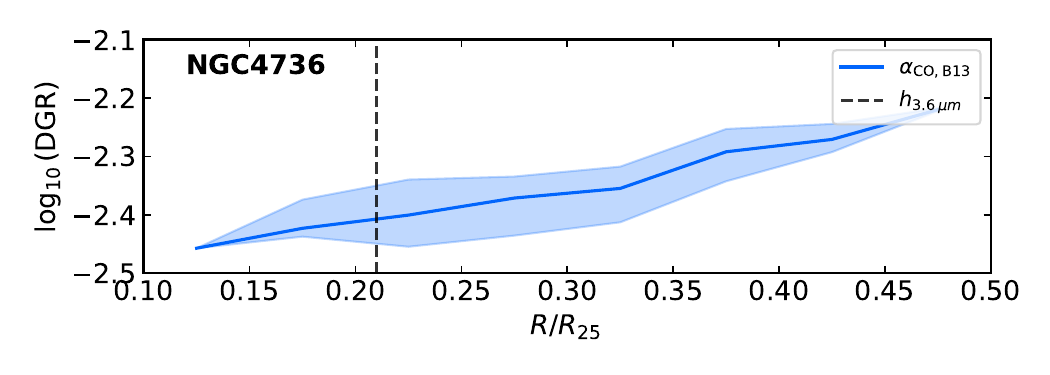} \\[1ex]
\includegraphics[width=0.49\textwidth]{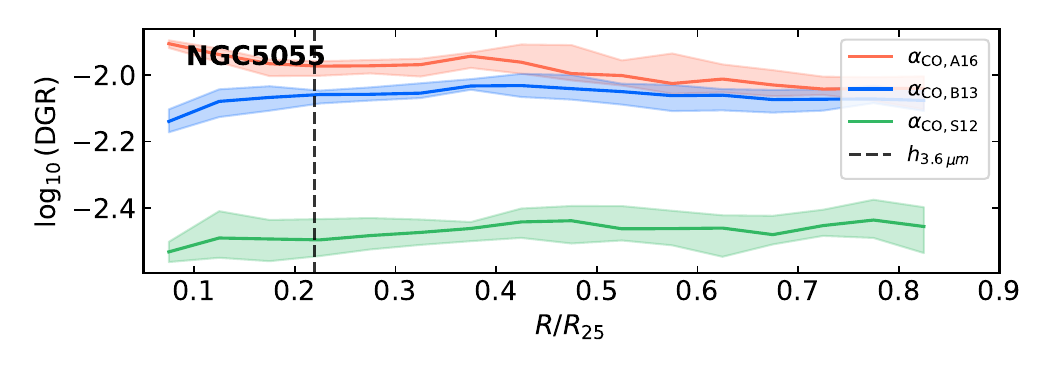} &
\includegraphics[width=0.49\textwidth]{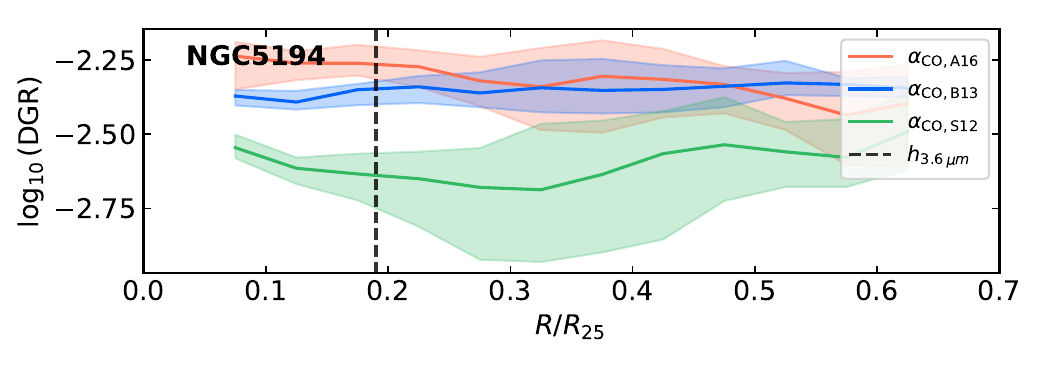}\\[1ex]
\includegraphics[width=0.49\textwidth]{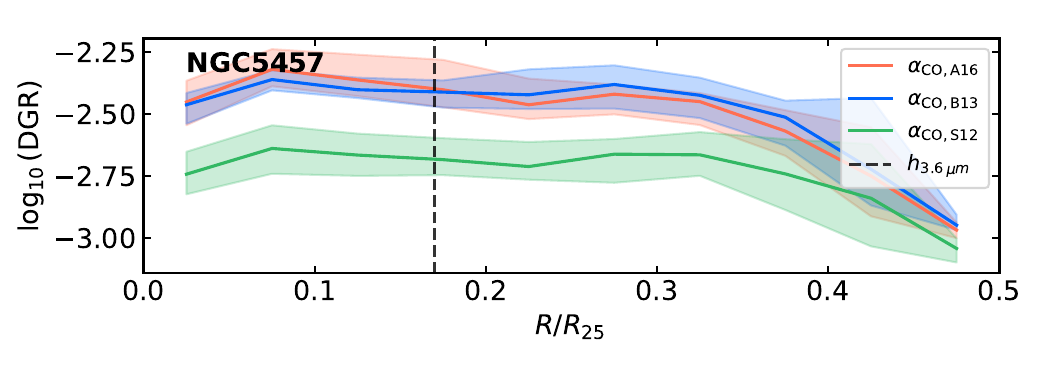}&
\includegraphics[width=0.49\textwidth]{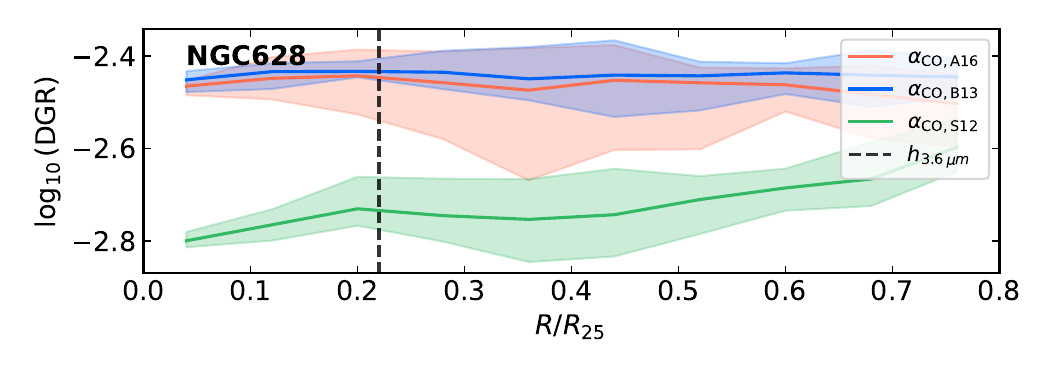} \\[1ex]
\includegraphics[width=0.49\textwidth]{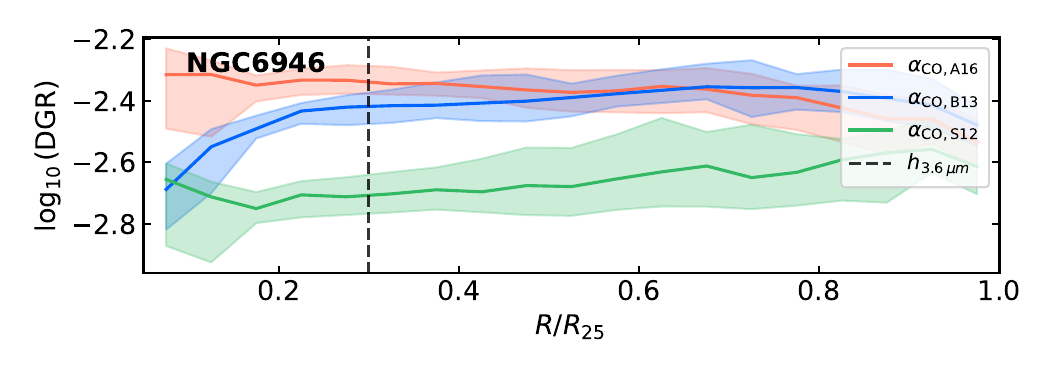} &
\includegraphics[width=0.49\textwidth]{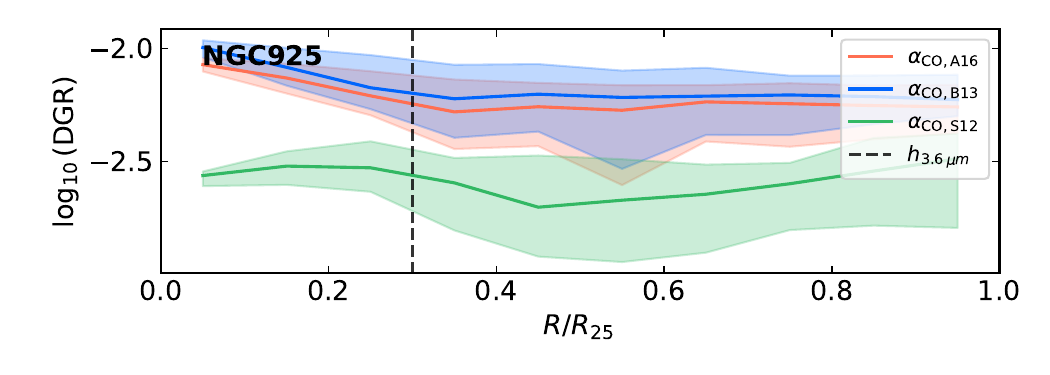} \\[1ex]
\end{tabular}
\caption{
Radial profiles of $\log(\mathrm{DGR})$
as a function of the normalized galactocentric radius ($R/R_{25}$) for the galaxy sample. Profiles are computed using three different $\alpha_{\rm CO}$ prescriptions: $\alpha_{\rm CO,B13}$ (blue), $\alpha_{\rm CO,A16}$ (red) and $\alpha_{\rm CO,S12}$ (green). Solid lines show the median values in radial bins, while shaded regions indicate the 16th–84th percentile range. Only regions with reliable detections in $\Sigma_{\rm HI}$, $\Sigma_{\rm H_2}$, $\Sigma_{\rm dust}$ and $\rm 12+log(O/H)$ are included. The vertical dashed line indicates the exponential disc scale length $h_{3.6\,\mu{\rm m}}$ from C17, which is consistent with the $K_s$-band effective radii derived from multi-component structural decompositions (Valerdi et al. 2026, in prep.; see also \cite{Rios_2025}), and is used here as an alternative physically motivated size scale that better traces the stellar mass distribution \citep{Baes_2024}. 
}
\end{figure}
\clearpage
\section{Radial Profiles for DMR}
\label{appendix: radial prof dmr}

\begin{figure}[htbp]
\centering
\begin{tabular}{cc}
\includegraphics[width=0.49\textwidth]{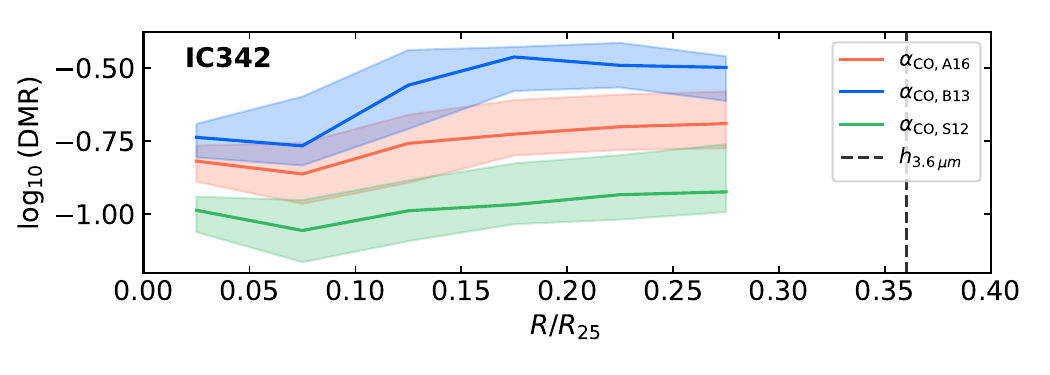} &
\includegraphics[width=0.49\textwidth]{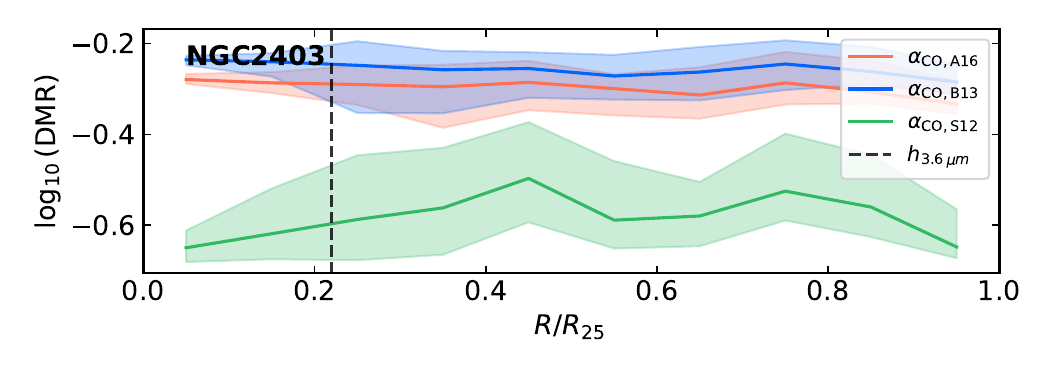} \\[1ex]
\includegraphics[width=0.49\textwidth]{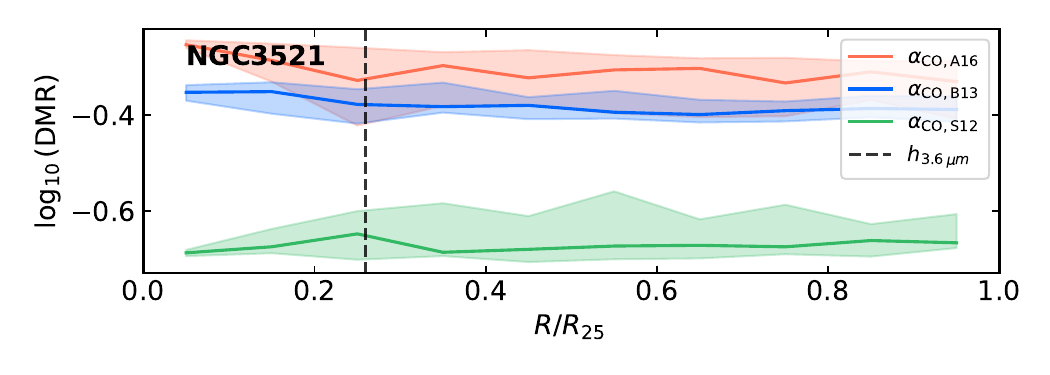} &
\includegraphics[width=0.49\textwidth]{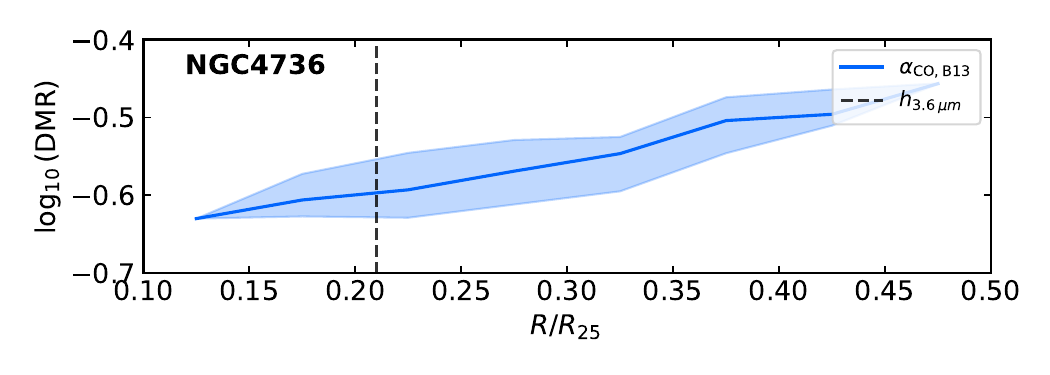} \\[1ex]
\includegraphics[width=0.49\textwidth]{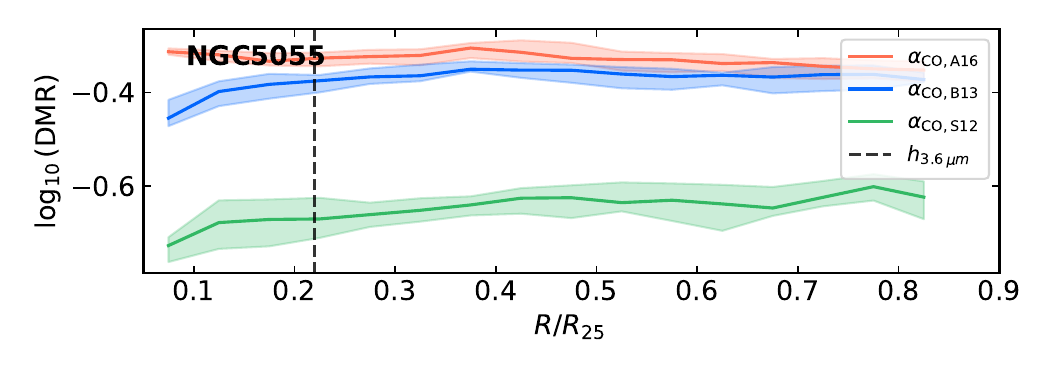} &
\includegraphics[width=0.49\textwidth]{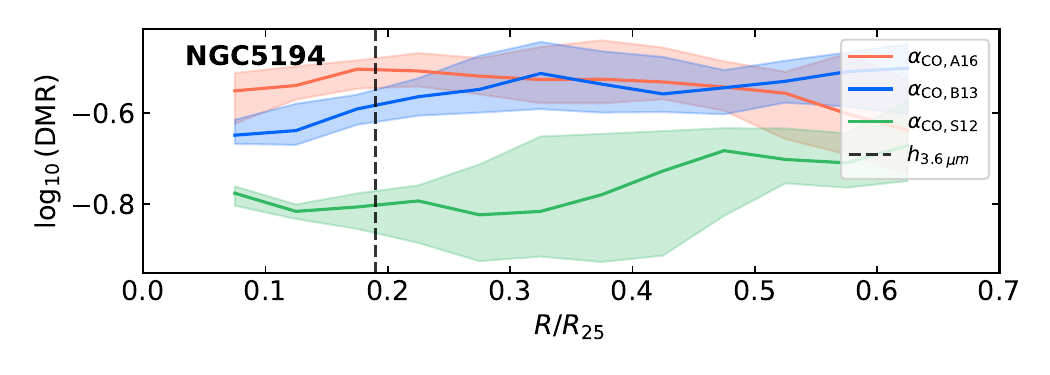}\\[1ex]
\includegraphics[width=0.49\textwidth]{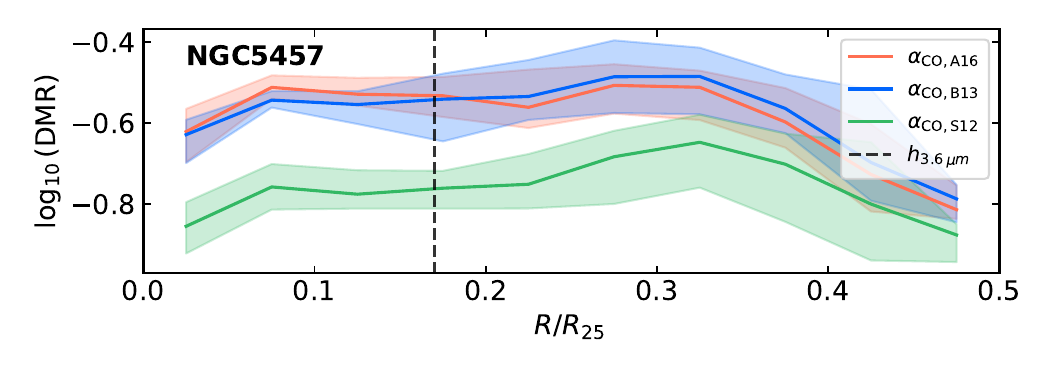}&
\includegraphics[width=0.49\textwidth]{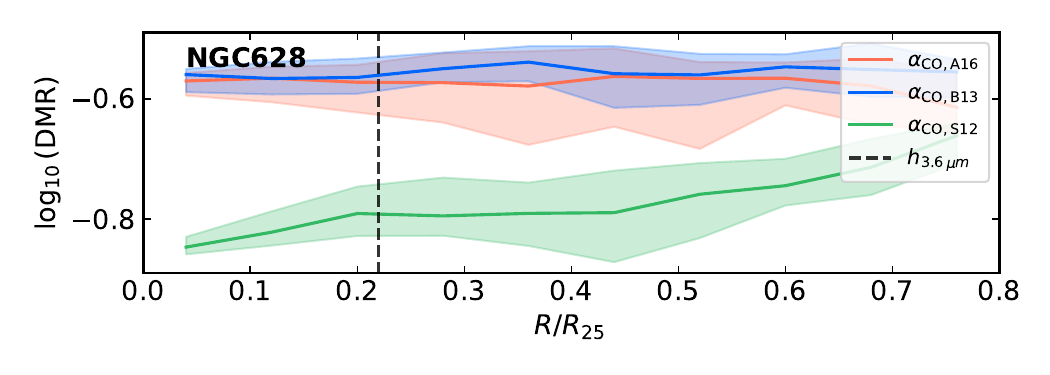} \\[1ex]
\includegraphics[width=0.49\textwidth]{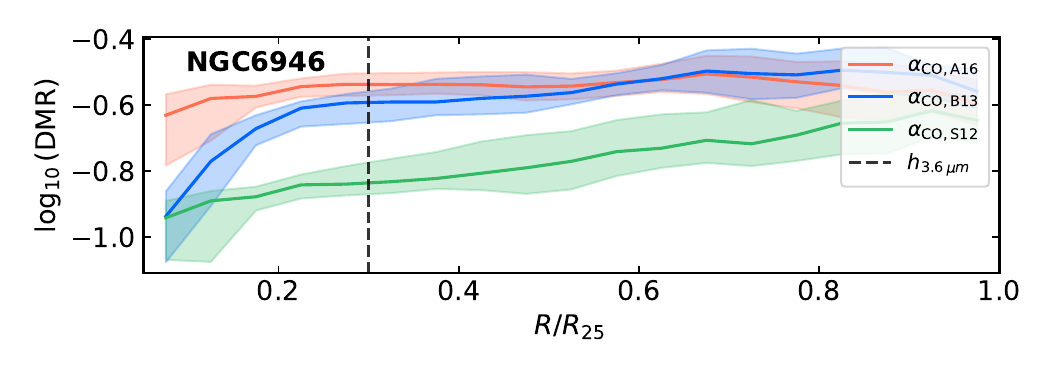} &
\includegraphics[width=0.49\textwidth]{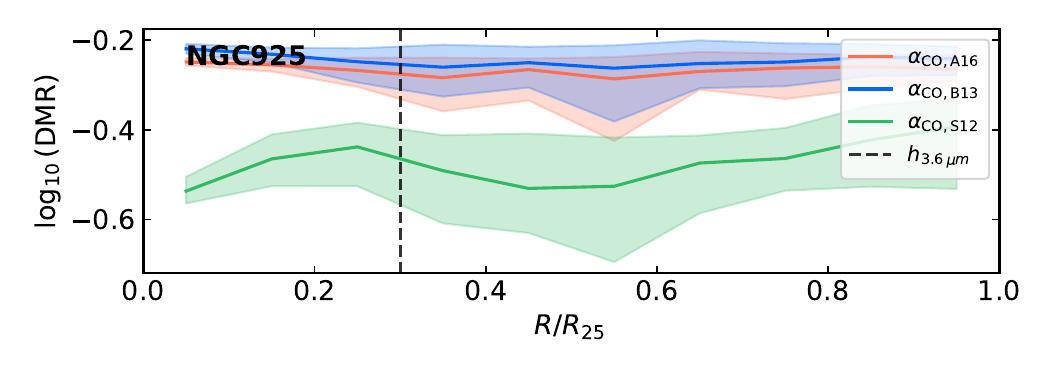} \\[1ex]
\end{tabular}
    \caption{
    Same as Fig.~\ref{appendix: radial prof dgr}, but showing the resolved $\log(\mathrm{DMR})$ profiles.
    }
\end{figure}
\end{appendix}
\end{document}